\documentclass[aps,pra,twocolumn]{revtex4-2}
\usepackage{graphicx,subfigure,epsfig,appendix}
\usepackage{booktabs,multirow,mathrsfs,amssymb,amsmath,amsthm}
\usepackage{algorithm,algorithmicx,framed,listings} 
\usepackage[colorlinks=true,linkcolor=blue,citecolor=blue,urlcolor=blue]{hyperref}
\usepackage[noend]{algpseudocode}
\usepackage{pifont}

\newcommand{\bra}[1]{\langle #1|}
\newcommand{\ket}[1]{|#1\rangle}
\newcommand{\inner}[2]{\langle #1|#2\rangle}
\graphicspath{{figures/}}
\begin{document}


\title{A Time-Symmetric Quantum Algorithm for Direct Eigenstate Determination}
\author{Shijie Wei$^{1}$}
\thanks{These authors contributed equally to this work.}
\author{Jingwei Wen$^{2}$}
\thanks{These authors contributed equally to this work.}
\author{Xiaogang Li$^{4,5}$}
\thanks{These authors contributed equally to this work.}

\author{Peijie Chang$^{3}$}

\author{Bozhi Wang$^{1,3}$}
\email{wang\_bozhi23@163.com}

\author{Franco Nori$^{6,7}$}
\email{fnori@riken.jp}
\author{Guilu Long$^{1,3,8,9}$}
\email{gllong@tsinghua.edu.cn}

\affiliation{$^{1}$ Beijing Academy of Quantum Information Sciences, Beijing 100193, China}
\affiliation{$^{2}$ China Mobile (Suzhou) Software Technology Company Limited, Suzhou 215163, China}
\affiliation{$^{3}$ State Key Laboratory of Low-Dimensional Quantum Physics and Department of Physics, Tsinghua University, Beijing 100084, China}
\affiliation{$^{4}$ Center on Frontiers of Computing Studies, Peking University, Beijing 100871, China}
\affiliation{$^{5}$ School of Computer Science, Peking University, Beijing 100871, China}

\affiliation{$^{6}$Center for Quantum Computing, RIKEN, Wakoshi, Saitama, 351-0198, Japan}

\affiliation{$^{7}$Quantum Research Institute, The University of Michigan, Ann Arbor, 48109-1040, MI, USA}
\affiliation{$^{8}$ Frontier Science Center for Quantum Information, Beijing 100084, China}
\affiliation{$^{9}$ Beijing National Research Center for Information Science and Technology, Beijing 100084, China}

\date{\today}


\begin{abstract}
Time symmetry in quantum mechanics, where the current quantum state is determined jointly by both the past and the future, offers a more comprehensive description of physical phenomena. This symmetry facilitates both forward and backward time evolution, providing a computational advantage over methods that rely on a fixed time direction.
In this work, we present a nonvariational and \textit{time-symmetric quantum algorithm} for addressing the eigenvalue problem of the Hamiltonian, leveraging the coherence between forward and backward time evolution. Our approach enables the simultaneous determination of both the ground state and the highest excited state, as well as the direct identification of arbitrary eigenstates of the Hamiltonian. Unlike existing methods, our algorithm eliminates the need for prior computation of lower eigenstates, allowing for the direct extraction of any eigenstate and  energy bandwidth while avoiding error accumulation. Its non-variational nature ensures convergence to target states without encountering the barren plateau problem.
We demonstrate the feasibility of implementing the non-unitary evolution using both the linear combination of unitaries and quantum Monte Carlo methods. Our algorithm is applied to compute the energy bandwidth and spectrum of various molecular systems, as well as to identify topological states in condensed matter systems, including the Kane-Mele model and the Su-Schrieffer-Heeger model. We anticipate that this algorithm will provide an efficient solution for eigenvalue problems, particularly in distinguishing quantum phases and calculating energy bands.

\end{abstract}
\maketitle


\section{Introduction}

Traditional quantum mechanics is not time-symmetric, as evolution always proceeds forward in time, with the current quantum state uniquely determined by the past. In 1964, the ABL (Aharonov, Bergmann, Lebowitz) theory introduced the concept of backward-evolving states and proposed the two-state vector formalism~\cite{Aharonov1964,Aharonov2010time,Kofman2012nonperturbative}, offering a novel perspective on quantum mechanics. In this framework, quantum mechanics becomes time-symmetric, with the current quantum state determined jointly by both the past and the future, enabling both forward and backward time evolution. This theory has inspired extensive research, significantly advancing our understanding of the quantum realm; and it has also sparked ongoing research into weak measurements~\cite{Aharonov1988, Kedem2010,Dressel2014, Imoto2018}. Furthermore, this idea has a significant impact on quantum simulation techniques~\cite{ Buluta2009quantum,RevModPhys.86.153}, potentially simplifying tasks that are difficult in traditional quantum mechanics. For example, non-Hermitian systems, which follow nonunitary evolution, such as parity-time symmetric systems~\cite{Huang2019}, can be conveniently simulated, and certain non-Hermitian observables, such as creation and annihilation operators, are easier to measure~\cite{Lundeen2005,Li2020}.

The forward and backward time evolution also finds applications in error mitigation~\cite{Filippov2023scalable} and beyond the Heisenberg uncertainty relation~\cite{Bao2020retrodiction}, through retrodiction, which involves interpreting past events by inference based on currently available information. Importantly, by using ancillary qubits, quantum measurements, and postselection techniques, forward and backward time evolutions can be transformed from a sequential evolution pattern into a parallel interference pattern. This transformation serves as a witness of time’s arrows~\cite{rubino2021quantum,Stromberg2024experimental} and provides a method for implementing quantum imaginary time evolution (QITE)~\cite{Kosugi2022}. Such a transformation changes the mathematical mechanism between evolution operators from multiplication to addition, surpassing the traditional unitary constraints.

In the quantum physics, the eigenvalue problem of the Hamiltonian is crucial because it determines the system's energy spectra and quantum states, providing the foundation for understanding particle behavior, dynamics, and interactions. There are various quantum algorithms that have been developed to calculate ground states, including quantum phase estimation (QPE)~\cite{abrams1997simulation,aspuru2005simulated,mcardle2020quantum}, variational quantum eigensolver (VQE)~\cite{peruzzo2014variational}, full quantum eigensolver (FQE)~\cite{wei2020full}, QITE~\cite{yeter2020practical,motta2020determining,sun2021quantum,wen2023iteration} and quantum machine learning (QML)~\cite{biamonte2017quantum,xia2018quantum,yoshioka2021solving}. 
Furthermore, understanding the excited states is equally crucial for predicting molecular behavior and reaction outcomes. Calculating the energy spectrum distribution can reveal additional properties of complex systems, such as the classification of metals and insulators, as well as topological phase transitions. 

Advanced quantum algorithms~\cite{mcclean2017hybrid,colless2018computation,jones2019variational,higgott2019variational,nakanishi2019subspace,parrish2019quantum,wen2021variational,zhang2021adaptive,bharti2022noisy,huang2023near,wang2024improving,wen2024full} are pivotal in this area, providing powerful tools for studying excited states and thereby driving progress in quantum chemistry and physics. Foundational algorithms such as variational quantum deflation (VQD)~\cite{jones2019variational,higgott2019variational,wen2021variational} have been adapted to find excited states by applying state-specific penalization terms. Additionally, the subspace search variational quantum eigensolver (SSVQE)~\cite{nakanishi2019subspace,parrish2019quantum} optimizes a superposition of orthogonal quantum states within a specific subspace to identify excited states. Meanwhile, the quantum subspace expansion (QSE)~\cite{mcclean2017hybrid, colless2018computation} extends the VQE to compute excited states more efficiently. A more recent development is the full quantum excited state solver (FQESS) algorithm~\cite{wen2024full}, a nonvariational approach that aims to obtain the excited state spectrum of a quantum chemistry Hamiltonian, offering advantages like faster convergence and robustness against noise without the need for classical optimization. Furthermore, the powered-FQE algorithm~\cite{wang2024improving} significantly reduces the actual number of runs in the FQE and FQESS by substituting the original operator with its powers, thus mitigating the exponential decay of success probability with increasing iterations, while also not requiring too many ancillary qubits.  

However, those quantum algorithms for excited states still face several challenges. On the one hand, variational methods like VQD, SSVQE and QSE often struggle with optimization processes, making it difficult to effectively converge to the target state within a large parameter space~\cite{bittel2021training}. On the other hand, methods such as FQESS, powered-FQE and VQD typically require pre-determination of lower energy states, which can lead to error accumulation and reduced fidelity when calculating higher excited states~\cite{vdurivska2025quantum}. These challenges highlight the complexities involved in implementing excited state quantum algorithms and the ongoing need for further research and innovation in this field.

In this work, we propose a nonvariational, \textit{time-symmetric quantum eigensolver} (TSQES) that addresses the challenges outlined above and introduces new capabilities by utilizing the superposition of forward and backward time evolution. Our algorithm has two main advantages: first, it can simultaneously find both the nearest and farthest eigenstates from a preset energy value, which means that we can determine the ground state and the highest excited state of the Hamiltonian, thus providing the energy bandwidth. The second one is that it can function as a \textit{quantum excited state direct solver} (QESDS), enabling the direct calculation of arbitrary eigenstates without the need for pre-determining lower eigenstates, thereby avoiding the error accumulation phenomenon, which is typically associated with the calculation error of lower excited states.

We begin by constructing a full quantum circuit implementation using the linear combination of unitaries (LCU) framework~\cite{gui2006general,childs2012hamiltonian}, and then enhance the algorithm into a non-iterative version by incorporating ancillary qubits. This modification reduces the resource cost associated with repeated quantum state measurements. Furthermore, we introduce the quantum Monte Carlo (QMC) method~\cite{Carlson2015,Zhang2022,Huggins2022,Koczor2024} to implement TSQES, effectively mitigating the issue of exponential decay in the success probability, commonly encountered in LCU-based quantum algorithms. The complexities of both algorithm frameworks are thoroughly analyzed. 

We demonstrate the feasibility and efficiency of the algorithm in computing the eigenstates of various molecular systems, highlighting its advantage in avoiding error accumulation in comparison to previous methods. Furthermore, we explore the impact of the evolution time parameter on performance. Moreover, the algorithm is applied to identify topological states in condensed matter systems, focusing on two significant models: the Kane-Mele model and the Su-Schrieffer-Heeger model. These applications underscore the advantages of our algorithm in directly solving specific excited states, thereby enabling the direct study of particular physical phenomena of interest.


\section{Time-Symmetric Quantum Eigensolver} \label{Chaptheory}

\begin{figure*}
    \centering
\includegraphics[width=0.95\linewidth]{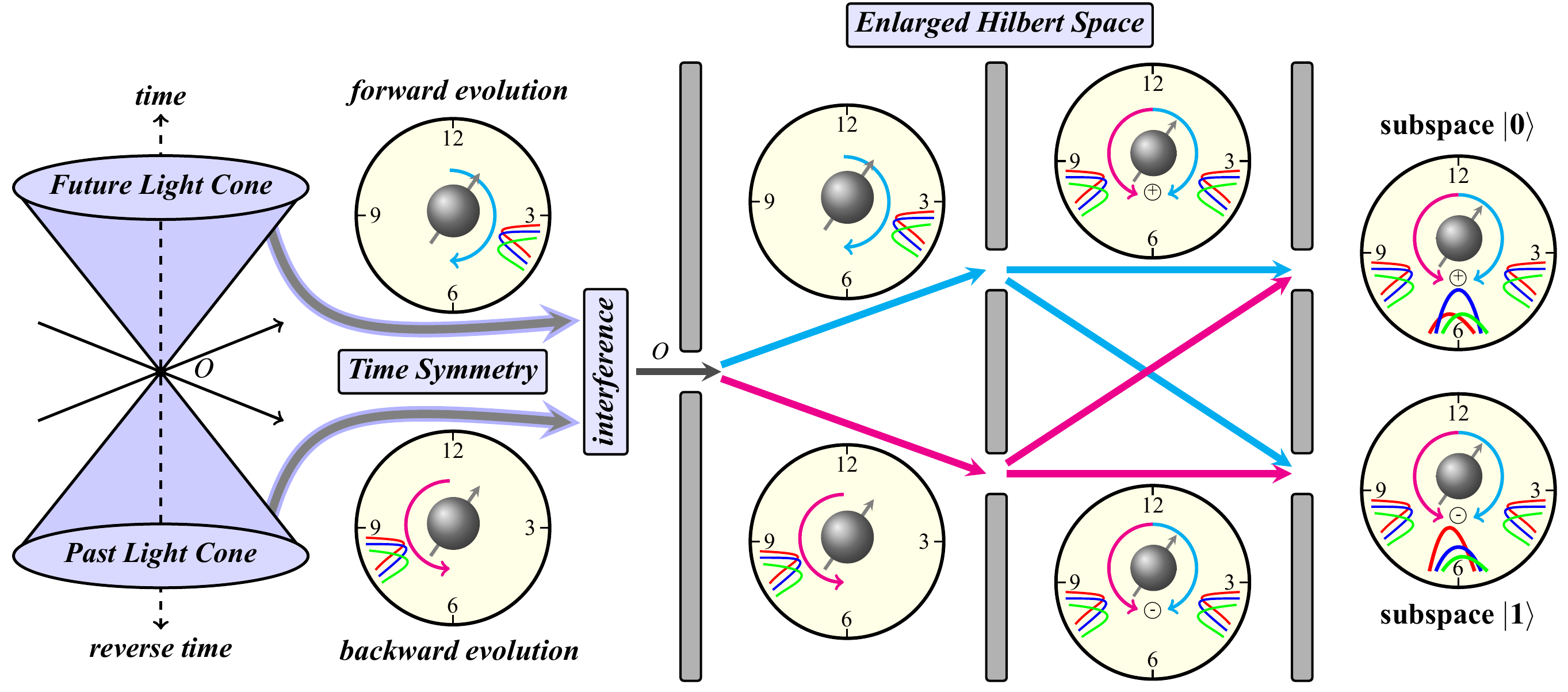}
       \caption{Schematic diagram of the time-symmetric quantum eigensolver. A quantum system evolves forward in time (clockwise) and backward in time (counterclockwise) and the interference will increase the amplitude of a specific eigenstate based on the choice of parameter setup in different subspaces. The arcs in the clock represent the components of the wave function, while the arrows indicate the direction of time evolution.}
        \label{fig1_diagram} 
\end{figure*}

Can we solve practical problems using the parallel interference pattern of forward-backward evolution? A natural physical intuition suggests that the mutual interference of time-symmetric forward-backward evolution can cancel out the phase oscillations, making the overall evolution nonunitary. This results in the eigenstates either gaining or dissipating symmetry. More specifically, considering a quantum system evolving simultaneously forward and backward in time over a given period and under the combined influence of both the past and the future, its final state becomes extreme, with the components of the ground state or the highest excited state increasing, leading to an enhanced specific state for the entire system. In this work, we realize the above mechanism by expanding the Hilbert space to propose the TSQES algorithm, which can achieve different eigenstates extraction in different subspaces, as shown in Fig.\ref{fig1_diagram}.

Most properties of many-body systems can be determined via the eigenstates of the Hamiltonian. Generally, the Hamiltonian includes the kinetic energies of both nuclei and electrons, as well as the Coulomb interactions among them. In its first quantized form, the Hamiltonian is expressed as
\begin{equation}
\label{eq:firstHam}
\begin{aligned}
H &= - \sum_i \frac{\nabla_{R_i}^2 }{2M_i} - \sum_i \frac{\nabla_{r_i}^2}{2} - \sum_{i,j} \frac{Z_i}{|R_i - r_j|} \\
&+ \sum_{i, j > i} \frac{Z_i Z_j}{|R_i - R_j|} + \sum_{i, j>i} \frac{1}{|r_i - r_j|}.
\end{aligned} 
\end{equation}
Here, $R_i$, $Z_i$, $M_i$ and $r_i$ represent the positions, charges, masses of the nuclei, and the positions of the electrons, respectively. This expression, given in atomic units, encompasses the fundamental interactions in the system.

After second quantization, the Hamiltonian can be expressed in the particle representation form
\begin{equation}
\begin{aligned}\label{eq:ferm_hamiltonian}
H= \sum_{ij} h_{ij} a^{\dagger}_i a_j+ \frac{1}{2} \sum_{ijkl} h_{ijkl} a^{\dagger}_i a^{\dagger}_j a_k a_l + \cdots,
\end{aligned}
\end{equation}
where $a_i^\dagger$ and $a_i$ denote the creation and annihilation operators for a particle in orbital $i$, while $h_{ij}$ and $h_{ijkl}$ are the one-particle and two-particle integrals defined in a chosen basis.

To perform calculations on the quantum computers, we map the fermionic operators to qubit operators using methods such as the Bravyi-Kitaev transformation~\cite{bravyi2002fermionic, seeley2012bravyi} or the Jordan-Wigner transformation~\cite{batista2001generalized, aspuru2005simulated,sboychakov2015electronic}. This mapping yields the qubit Hamiltonian

\begin{equation}
\begin{aligned}\label{eq:qubit_hamiltonian}
H=&\sum_{i,\alpha}p_{\alpha}^i\sigma_{\alpha}^i+\sum_{i,j,\alpha,\beta}p_{\alpha\beta}^{ij}\sigma_{\alpha}^{i}\sigma_{\beta}^j+\cdots,
\end{aligned}
\end{equation}
where the subscript $i (j)$ denotes the qubit on which the operator acts, and the superscript $\alpha (\beta)$ refers to the type of Pauli operators. Apparently, $H$ in Eq.\eqref{eq:qubit_hamiltonian} is a linear combination of unitary Pauli terms and we discuss our quantum algorithms in such Hamiltonian forms below.


For an $n$-qubit quantum system, the eigen-equation for the Hamiltonian is expressed as $H\ket{E_{i}}=E_{i}\ket{E_{i}}$, where $\ket{E_{i}}$ are the eigenstates and $E_{i}$ are the corresponding eigenvalues, satisfying $E_{1}<E_{2}<\cdots<E_{2^n}$. Then any initial quantum state $\ket{\psi_{0}}$ can be expanded on the complete basis of eigenstates by
$\ket{\psi_{0}}=\sum_{i=1}^{2^n}a_{i}\ket{E_{i}}$,
where the complex coefficients satisfy the normalization condition $\sum_{i=1}^{2^n}\vert a_{i}\vert^2=1$. 

We construct a forward time evolution operator with an energy-shift $e_{s}$ as $U_{f}^{e_{s}}=\exp(-i(H-e_{s})t)$, and a corresponding backward time evolution operator as $U_{b}^{e_{s}}=\exp(i(H-e_{s})t)$, where $t$ represents the evolution time.
By expanding the Hilbert space, we can achieve both constructive and destructive interference of the operators $U_{f}^{e_{s}}$ and $U_{b}^{e_{s}}$ simultaneously. 
Performing  the combinations of these two operators by $|0\rangle \langle0|\otimes U_{f}^{e_{s}}$ and $|1\rangle \langle1|\otimes U_{b}^{e_{s}}$. After $k$ times iterations in subspace $m_a=|0\rangle $ and $m_a=|1\rangle$, the initial state $\ket{\psi_{0}}$ is transformed to 
\begin{equation}
\begin{split}
\ket{\psi_{k}}_0&=(U_{f}^{e_{s}}+U_{b}^{e_{s}})^{k}\ket{\psi_{0}}\\
&=\sum_{i=1}^{2^n}a_{i}\big[2\cos((E_{i}-e_{s})t) \big]^{k}\ket{E_{i}}\\
&=\big[2\cos((E_{m}-e_{s})t) \big]^{k}\Big[ a_{m}\ket{E_{m}}+\\
&\sum_{i\ne m}^{2^n}a_{i}\Big(\frac{\cos((E_{i}-e_{s})t)}{\cos((E_{m}-e_{s})t)}\Big)^{k}\ket{E_{i}} \Big].
\end{split}
\label{psi_k1}
\end{equation}

and 
{\begin{equation}
    \begin{split}
    \ket{\psi_{k}}_1&=(U_{f}^{e_{s}}-U_{b}^{e_{s}})^{k}\ket{\psi_{0}}\\
    &=\sum_{i=1}^{2^n}a_{i}\big[-2i\sin((E_{i}-e_{s})t) \big]^{k}\ket{E_{i}}\\
    &=\big[-2i\sin((E_{\mathrm{max}}-e_{s})t) \big]^{k}\Big[ a_{\mathrm{max}}\ket{E_{\mathrm{max}}}+\\
&\sum_{i\ne \mathrm{max}}^{2^n}a_{i}\Big(\frac{\sin((E_{i}-e_{s})t)}{\sin((E_{\mathrm{max}}-e_{s})t)}\Big)^{k}\ket{E_{i}} \Big].
    \end{split}
    \label{psi_k2}
    \end{equation}
respectively. If we set $(E_{i}-e_{s})t \in [-\pi/2,\pi/2]~\forall~i$, then the $m$-th eigenstate $\ket{E_{m}}$ with eigenvalue $E_{m}$ closest to $e_{s}$ in subspace $|0\rangle $ will have the maximum amplitude in $\ket{\psi_{k}}_0$ as $\vert a_m[2\cos((E_{m}-e_{s})t)]^k \vert$. And in subspace $|1\rangle $, the $\mathrm{max}$-th eigenstate $\ket{E_{\mathrm{max}}}$ with eigenvalue $E_\mathrm{{max}}$ farthest to $e_{s}$ will have the maximum amplitude in $\ket{\psi_{k}}_1$ as $\vert a_{\mathrm{max}}[2\sin((E_{\mathrm{max}}-e_{s})t)]^k \vert$. As $k$ increases, the contribution of the non-target eigenstates diminishes, and interference will increase the amplitude of the specific eigenstate. 

Upon normalization, the state $\ket{E_{m}}$ or $\ket{E_{\mathrm{max}}}$ can be approximately and directly obtained. If $e_{s}$ is close to the ground state energy, $\ket{E_{m}}$ is equal to the ground state $ \ket{E_{0}}$ and  $\ket{E_{\mathrm{max}}}$ is equal to the highest exited state. Then, the energy bandwidth can be obtained. When we only focus on the subspace $|0\rangle $, the  TSQES is reduced to a QESDS algorithm, which can directly determine one specific eigenstate $\ket{E_{m}}$.

However, the process of Eq.(\ref{psi_k1}) and Eq.(\ref{psi_k2}) is non-unitary and cannot be implemented directly with quantum circuits. Below we present two strategies using LCU and QMC methods for the realization of this TSQES algorithm.


\begin{figure}
    \centering
\includegraphics[width=0.95\linewidth]{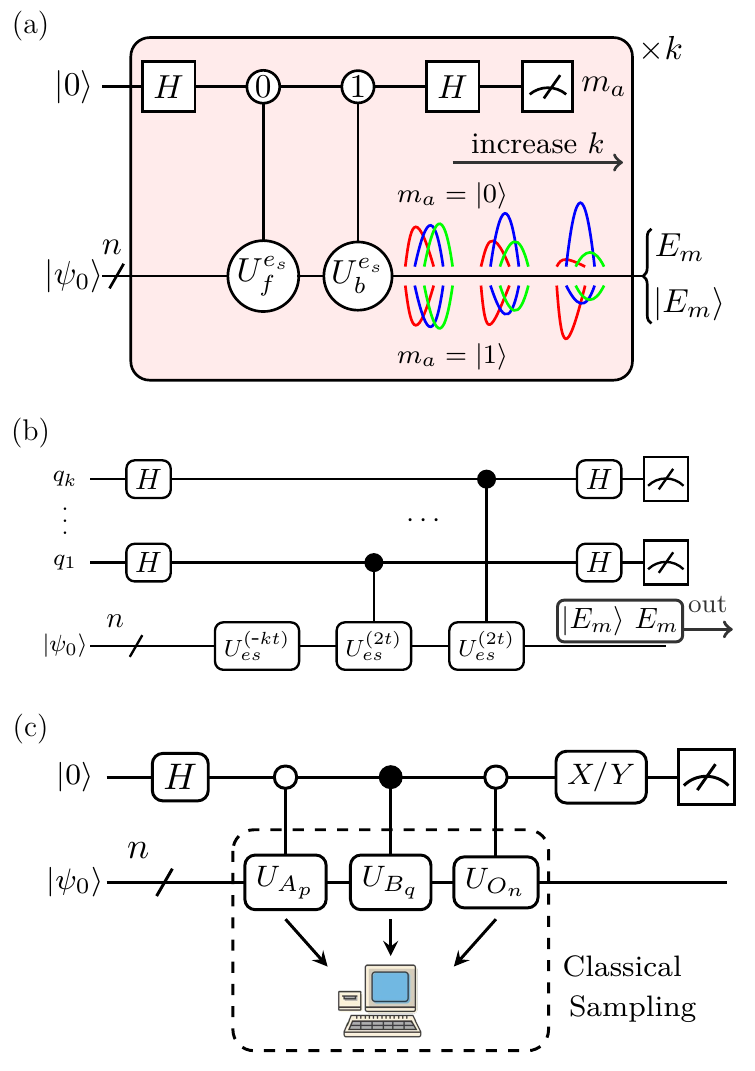}
       \caption{ Quantum circuits for implementing time-symmetric quantum eigensolver. (a) Quantum circuit with one ancillary qubit for the realization of superposition of the forward and backward time evolution. (b) Quantum circuit for the realization of the order-$k$ iteration-free  TSQES algorithm. (c) The principle for the QMC algorithm to compute the target values $\langle\psi|B^\dag O A|\psi\rangle$, where the observable $O=\sum_n c_n U_{O_n}$, $A=\sum_p a_p U_{A_p}$, and $B=\sum_q b_q U_{B_q}$. The samples $(p,q,n)$ are generated on a classical computer according to the relevant probability distribution. The $X$ (or $Y$) at the end of the circuit represents the measurement basis selected, which is respectively related to the real (or imaginary) parts of the target value.}
        \label{fig1_circuit} 
\end{figure}

\section{Time-Symmetric Quantum Eigensolver in the LCU frame}

Here we describe how to perform the combination of forward and backward time evolution simultaneously on a quantum computer in the LCU frame, and the quantum circuit for the realization of the TSQES algorithm is shown in Fig.\ref{fig1_circuit}\textcolor{blue}{(a)}. Consider a composite system consisting of a working system and an ancillary qubit with initial state $|0\rangle|\psi_0 \rangle$, we implement a Hadamard gate on the ancillary qubit first and then perform the ancillary-qubit controlled gates $ |0\rangle \langle 0|\otimes \exp(-i(H-e_s)t) $ and  $ |1\rangle \langle 1|\otimes \exp(i(H-e_s)t) $ on the whole system. Inverting the ancillary qubit by a Hadamard gate the evolution becomes
\begin{equation}
        \begin{split}
        &\frac{1}{2}|0\rangle [\exp(-i(H-e_s)t)+\exp(i(H-e_s)t)]|\psi_0 \rangle+ \\
        &\frac{1}{2}|1\rangle [\exp(-i(H-e_s)t)-\exp(i(H-e_s)t)]|\psi_0 \rangle.
        \end{split}
\end{equation}

If the ancillary qubit collapses into the $|0\rangle\bra{0}$ state, with probability $\|\cos((H-e_s)t)\|^2$, the working system evolves as $| \psi_1 \rangle=[\exp(-i(H-e_s)t)+\exp(i(H-e_s)t) ]|\psi_0\rangle$. Otherwise, the working system undergoes an evolution $[\exp(-i(H-e_s)t)-\exp(i(H-e_s)t)]|\psi_0 \rangle$ with probability $\|\sin((H-e_s)t)\|^2$. We continue to perform the same evolution process when the ancillary qubit is measured as $|0\rangle$ or $|1\rangle$; and after $k$-th iterations, the working system will evolve to the quantum state $\ket{\psi_{k}}_0$ or $\ket{\psi_{k}}_1$, which is approximate to the eigenstate $|E_m \rangle$ or $|E_{\mathrm{max}} \rangle$. 

In practice, the forward and backward time evolution operators are repeatedly applied on the quantum state until the energy measurements $\bra{\psi_k}H\ket{\psi_k}$ stabilize. The output state is the specific eigenstate of the Hamiltonian, whose corresponding eigenvalue is most adjacent or distant to the energy-shift $e_{s}$. It should be noted that the TSQES algorithm requires pre-knowledge of the target eigenenergy value, which guides the choice of $e_{s}$, and the initial state needs to have a limited overlap with the target eigenstate.

In addition to using the iteration process to realize the increase of order-$k$, we can extend the TSQES  algorithm to an iteration-free version by introducing more ancillary qubits to eliminate the iteration and measurement times, as shown in Fig.\ref{fig1_circuit}\textcolor{blue}{(b)}. For an order-$k$ extension, the algorithm requires $(n+k)$ qubits. Specifically, we can mathematically express the superposition of forward and backward time evolution operators as $U_{f}^{e_{s}}\pm U_{b}^{e_{s}} =\exp(-i(H-e_s)t)[1\pm \exp(2i(H-e_s)t)]$. If we formally introduce the operator $U_{es}^{(t)}=\exp(i(H-e_s)t)$ to unify the representation of the two operators, then the order-$k$ evolution will be 
\begin{equation}
        \begin{split}
(U_{f}^{e_{s}}\pm U_{b}^{e_{s}})^k=U_{es}^{(-kt)}(1\pm U_{es}^{(2t)})^k.
        \end{split}
\end{equation}
where the first part $U_{es}^{(-kt)}$ is a forward time evolution operator and can be applied directly to the working system, while the realization of the second part needs the LCU frame by using ancillary qubits. After encoding the ancillary system with Hadamard gates, a series of ancillary qubits $q_{i}~(i=1,2,\cdots,k)$ controlled operators are applied on  the working system as $\ket{0}\bra{0}_{i}\otimes I^{\otimes n}+\ket{1}\bra{1}_{i}\otimes U_{es}^{(2t)}$. 

In order to guarantee the success of the algorithm, the condition $\vert(E_{i}-e_{s})t\vert <\pi/2$ still needs to be satisfied. Then we decode the system by adding Hadamard gates on each ancillary qubit, which is the inverse action of the encoding process. At the end of the circuit, projective measurements are applied to the ancillary qubits. In the $\ket{0}^{\otimes k}$ subspace, we can obtain the quantum state $\ket{E_m}$ in the working system, whose eigenvalue $E_{m}=\bra{E_m}H\ket{E_m}$ is most adjacent to $e_{s}$. In the $\ket{1}^{\otimes k}$ subspace, we can obtain the quantum state $\ket{E_{\mathrm{max}}}$ in the working system, whose eigenvalue $E_{\mathrm{max}}=\bra{E_{\mathrm{max}}}H\ket{E_{\mathrm{max}}}$ is farthest from  $e_{s}$. The output can be the ground (highest exited) state or an arbitrary excited state (the state farthest from the specific excited state) depending on the values of the energy-shift parameter $e_s$.

Compared to the original method, the iteration-free method no longer requires $k$ repetitions of the quantum circuits. We only need to perform the quantum circuit once, and the number of controlled evolution gates can be reduced by half at the cost of additional $(k-1)$ ancillary qubits. These two methods can be regarded as a trade-off between quantum resources (qubit number and circuit depth) and times of measurement. In practice, we can choose a compromise strategy according to the actual quantum hardware requirements; that is, using $m$ ancillary qubits and $k/m$ iterations to complete the order-$k$ calculation.


Next, we analyze the complexity of the TSQES algorithm in the LCU-type frame. The normalized quantum state after an order-$k$ quantum circuit can be expressed as 
\begin{equation}
\ket{\psi_{k}}_j=\sum_{i=1}^{2^n}a_{i}\lambda_{ij}^{k}\ket{E_{i}}/\sqrt{C},
\end{equation}
with $\lambda_{i0}=2\cos((E_{i}-e_{s})t)$, $\lambda_{i1}=2i\sin((E_{i}-e_{s})t)$ and $C=\sum_{i=1}^{2^n}\vert a_{i}\vert^{2}\lambda_{ij}^{2k}$, $j=0,1$. We assume $|\lambda_{sj}|>|\lambda_{tj}|\ge\cdots$, and they represent the two eigen-components that are most adjacent (distant) to $e_s$. The probability of finding the system in the eigenstate $\ket{E_{s}}$ is
\begin{equation}
\begin{split}
P(\ket{E_{s}})=&\vert\inner{E_{s}}{\psi_{k}}_j\vert^{2}=\frac{\vert a_{s}\vert^{2}\lambda_{sj}^{2k}}{\sum_{i=1}^{2^n}\vert a_{i}\vert^{2}\lambda_{ij}^{2k}}\\
&=\frac{\vert a_{s}\vert^{2}}{\vert a_{s}\vert^{2}+\sum_{i\ne s}^{2^n}\vert a_{i}\vert^{2}(\lambda_{ij}/\lambda_{sj})^{2k}}\\
&\geqslant\frac{\vert a_{s}\vert^{2}}{\vert a_{s}\vert^{2}+\sum_{i\ne s}^{2^n}\vert a_{i}\vert^{2}(\lambda_{tj}/\lambda_{sj})^{2k}}\\
&\propto \left(\frac{\lambda_{sj}}{\lambda_{tj}}\right)^{2k},
\end{split}
\label{eqbound}
\end{equation}
with 
\begin{equation}
\begin{split}
\left(\frac{\lambda_{s0}}{\lambda_{t0}} \right)^{2k} =\left[\frac{\cos((E_{s}-e_{s})t)}{\cos((E_{t}-e_{s})t)}\right]^{2k},\\
\left(\frac{\lambda_{s1}}{\lambda_{t1}}\right)^{2k} =\left[\frac{\sin((E_{s}-e_{s})t)}{\sin((E_{t}-e_{s})t)}\right]^{2k}.
\end{split}
\end{equation}

Eq.(\ref{eqbound}) indicates that the lower bound of the overlap with the target quantum state increases exponentially with iterations $k$. A good choice of parameters $e_s$ and $t$ will accelerate the convergence and reduce the number of ancillary qubits in the iteration-free scheme. Usually, we can reasonably assume that the exact values of the energy $\{E_s\}$ are unknown, and we only know an approximate value calculated by the classical method. For a large value of the iteration $k$, the approximation error for the energy $E_{s}$ becomes

\begin{equation}
\begin{aligned}
\epsilon & =\left|\left\langle\psi_k|H| \psi_k\right\rangle_j-E_{s}\right| \\
& =\frac{\sum_{i=1}^{2^n}\left|a_i\right|^2 \lambda_{ij}^{2 k}\left|E_{i}-E_{s}\right|}{\sum_{i=1}^{2^n}\left|a_i\right|^2 \lambda_{ij}^{2 k}} \\
& =\frac{\sum_{i \neq s}^{2^n}\left|a_i\right|^2\left(\frac{\lambda_{ij}}{\lambda_{sj}}\right)^{2 k}\left|E_{i}-E_{s}\right|}{\left|a_{s}\right|^2+\sum_{i \neq s}^{2^n}\left|a_i\right|^2\left(\frac{\lambda_{ij}}{\lambda_{sj}}\right)^{2 k}} \\
& \leqslant \frac{\sum_{i \neq s}^{2^n}\left|a_i\right|^2\left(\frac{\lambda_{tj}}{\lambda_{sj}}\right)^{2 k}\left|E_{i}-E_{s}\right|}{\left|a_{s}\right|^2} \\
& \leqslant \frac{\|H\|_1}{|a_{s}|^2}\cdot\left(\frac{\lambda_{tj}}{\lambda_{sj}}\right)^{2k},
\end{aligned}
\end{equation}
where the error decreases exponentially with the iterations $k$. Then taking the logarithm of both sides of the above inequality, we can obtain 
\begin{equation}
\begin{aligned}\label{kk}
k\leqslant\frac{\log(\frac{\epsilon|a_{s}|^2}{\|H\|_1})}{2\log(\frac{\lambda_{tj}}{\lambda_{sj}})}=\frac{\log(\frac{\|H\|_1}{\epsilon|a_{s}|^2})}{2\log(\frac{\lambda_{sj}}{\lambda_{tj}})}=O\left(\log\frac{1}{\epsilon}\right).
\end{aligned}
\end{equation}

Notably, a good choice of parameters $e_s$  and evolution time $t$ is crucial to accelerating error reduction and improving the efficiency of the algorithm.


\section{Time-Symmetric Quantum Eigensolver with the QMC method}

We now address the exponential decay problem of the success probability with system dimension in the LCU-type quantum frame, which means that an additional exponentially deep amplitude amplification quantum circuit needs to be attached to amplify the post-selection probability. To do this, we alternatively adopt a QMC method to implement TSQES to alleviate this problem~\cite{Yang2021,huo2023error}, as shown in Fig.\ref{fig1_circuit}\textcolor{blue}{(c)}. The advantage of the QMC method is that it can reduce the number of auxiliary qubits, reduce the number of quantum  gates, especially the low fidelity two-qubit gates, and reduce the circuit depth, which is of great significance in the NISQ era. 

According to Eq.(\ref{psi_k1}) and Eq.(\ref{psi_k2}), the normalized quantum state after $k$ iterations is 
\begin{equation}\label{normalized_psik}
\begin{aligned}
\left|\psi_k\right\rangle= \frac{\left(U_f^{e_s}\pm U_b^{e_s}\right)^k \left|\psi_0\right\rangle}{\|\left(U_f^{e_s}\pm U_b^{e_s}\right)^k\left|\psi_0\right\rangle\|}
=\begin{cases}
\frac{\cos^k((H-e_s)t)|\psi_0\rangle}{D},&j=0 \\
\frac{(-i)^k\sin^k((H-e_s)t)|\psi_0\rangle}{D}, & j=1\\
\end{cases}
\end{aligned}
\end{equation}
where the normalized constant $D$  has an upper bound $\|((U_f^{e_s}\pm U_b^{e_s})/2)^k\left|\psi_0\right\rangle\|^2 \nonumber.$

Without loss of generality, we focus on the constructive interference of forward and backward time evolution corresponding to the time evolution in subspace $\ket{0}$. In this circumstance, TSQES is reduced to QESDS. The normalized constant becomes
\begin{equation}
\begin{aligned}\label{norm_denominator}
D&\equiv\left|\left|\left(\frac{U_f^{e_s}+U_b^{e_s}}{2}\right)^k\left|
\psi_0\right\rangle\right|\right|^2 \nonumber \\
&=\|\cos^k ((H-e_s)t)|\psi_0\rangle\|^2 \nonumber \\
&=\|\sum_{i} a_i\left[ \cos \left((E_i-e_s\right) t\right)]^k\left|E_i\right\rangle\|^2 \nonumber \\ 
&\geqslant \|a_s\left[\cos(\left(E_s-e_s\right) t\right)]^k\left|E_s\right\rangle\|^2 \nonumber \\
&=|a_s|^2\left[ \cos( \left(E_s-e_s\right) t\right)]^{2k},
\end{aligned}
\end{equation}
and in a similar way, we know that $|a_s|^2\lambda_s^{2k}\leqslant D\leqslant \lambda_s^{2k}$. Then we define $N(O)$ as
\begin{equation}
\begin{aligned}
N(O)\equiv&\langle\psi_0|\cos^k((H-e_s)t)O\cos^k((H-e_s)t)|\psi_0\rangle\\
=&\sum_n c_n\sum_{i,j} a_ia_j^* \left[\cos(\left(E_i-e_s\right) t\right)]^{k}\times\\
& \left[ \cos(\left(E_j-e_s\right) t)\right]^{k}\langle E_j|\sigma_n\left|E_i\right\rangle.
\end{aligned}
\end{equation}
Here the operator $O$ is decomposed into a linear combination of Pauli matrices $\sum_{n=1}^{G} c_n\sigma_n$. Then, we define the norm of $O$ as $\|O\|_1\equiv\sum_{n=1}^{G}|c_n|$, so $p_n=|c_n|/\|O\|_1$ is the probability of the occurrence of $\sigma_n$.  
 
Therefore, the expectation value of the observable $O$ can be expressed as
\begin{widetext}
\begin{equation}
\begin{aligned}\label{E}
\langle O\rangle&=\langle\psi_k|O|\psi_k\rangle=\frac{N(O)}{D} =\frac{\mathbb{E}_{k_1,k_2}\hat{N}(k_1,k_2)}{\mathbb{E}_{k_2^{'},k_1^{'}}\hat{D}(k_1^{'},k_2^{'}) } \\
&\equiv\frac{ \mathbb{E}_{k_1,k_2}\left[\langle\psi_0|({U_f^{e_s}}^\dag)^{2k_2-k}\cdot O \cdot (U_f^{e_s})^{2k_1-k}|\psi_0\rangle \right]}{\mathbb{E}_{k_1^{'},k_2^{'}}\left[\langle\psi_0|({U_f^{e_s}}^\dag)^{2k_2^{'}-k}\cdot (U_f^{e_s})^{2k_1^{'}-k}|\psi_0\rangle\right]} \\
&=\frac{ \mathbb{E}_{k_1,k_2}\left[\langle\psi_0|(U_f^{e_s})^{k-2k_2}\cdot O \cdot (U_f^{e_s})^{2k_1-k}|\psi_0\rangle \right]}{\mathbb{E}_{k_1^{'},k_2^{'}}\left[\langle\psi_0| (U_f^{e_s})^{2k_1^{'}-2k_2^{'}}|\psi_0\rangle\right]} \\
&=\frac{ \|O\|_1\cdot\sum p_n\mathbb{E}_{k_1,k_2}\left[\langle\psi_0|(U_f^{e_s})^{k-2k_2}\cdot \mathrm{sign}(c_n)\sigma_n \cdot (U_f^{e_s})^{2k_1-k}|\psi_0\rangle \right]}{\mathbb{E}_{k_1^{'},k_2^{'}}\left[\langle\psi_0| (U_f^{e_s})^{2k_1^{'}-2k_2^{'}}|\psi_0\rangle\right]} \\
&=\frac{ \|O\|_1\cdot\sum \mathbb{E}_{n,k_1,k_2}\left[\langle\psi_0|(U_f^{e_s})^{k-2k_2}\cdot \mathrm{sign}(c_n)\sigma_n \cdot (U_f^{e_s})^{2k_1-k}|\psi_0\rangle \right]}{\mathbb{E}_{k_1^{'},k_2^{'}}\left[\langle\psi_0| (U_f^{e_s})^{2k_1^{'}-2k_2^{'}}|\psi_0\rangle\right]}.
\end{aligned}
\end{equation}
\end{widetext}

Here, the numerator estimator is 
\begin{equation}
\begin{split}
\hat{N}(n,k_1,k_2)\equiv&\|O\|_1\langle\psi_0|(U_f^{e_s})^{k-2k_2}\cdot \mathrm{sign}(p_n)\sigma_n \cdot \\
&(U_f^{e_s})^{2k_1-k}|\psi_0\rangle \in[-\|O\|_1,\|O\|_1]
\end{split}
\end{equation}
and the denominator estimator is 
\begin{equation}
\hat{D}(k_1^{'},k_2^{'})\equiv\langle\psi_0| (U_f^{e_s})^{2k_1^{'}-2k_2^{'}}|\psi_0\rangle\in[-1,1]
\end{equation}
The symbol $\mathbb{E}$ represents the expectation about the random variables, such as \{$k_1,k_2$\} or \{$n,k_1,k_2$\}. Therefore, the QMC algorithm for denominator estimation can be efficiently implemented by computing the numerator $N(O)$, and the denominator $D$ separately. Specifically, when the observable $O$ is the Hamiltonian $H$, this method can be used to estimate the eigenvalue $E_j$ of $H$. The sampling complexity analysis of $D$ and $N(O)$, together with the quantum circuits, will be presented in Appendix \ref{appendix_A}.



In the following, we will first show the feasibility of the TSQES algorithm to simultaneously find both the ground state and the highest excited state of the Hamiltonian with the hydrogen molecule, thus providing the energy bandwidth. Then, we focus on the subspace $\ket{0}$, where the TSQES algorithm is reduced to the QESDS, and discuss the algorithm features including the error-accumulation phenomenon and the effect of the evolution time with a six-qubit LiH molecule. Finally, by solving the topological states in condensed matter systems, we show the advantages of the QESDS algorithm brought by the mechanism of directly finding specific eigenstates.


\section{Performance of the time-symmetric quantum eigensolver} \label{Chapsimulation}

\subsection{Calculating the energy bandwidth}

 \begin{figure}
    \centering
     \includegraphics[width=0.8\linewidth]{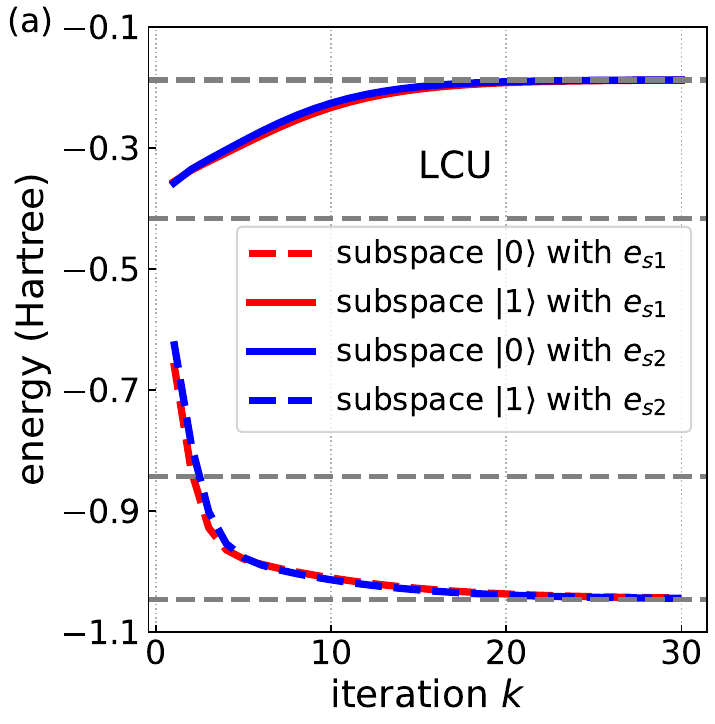}
     \includegraphics[width=0.8\linewidth]{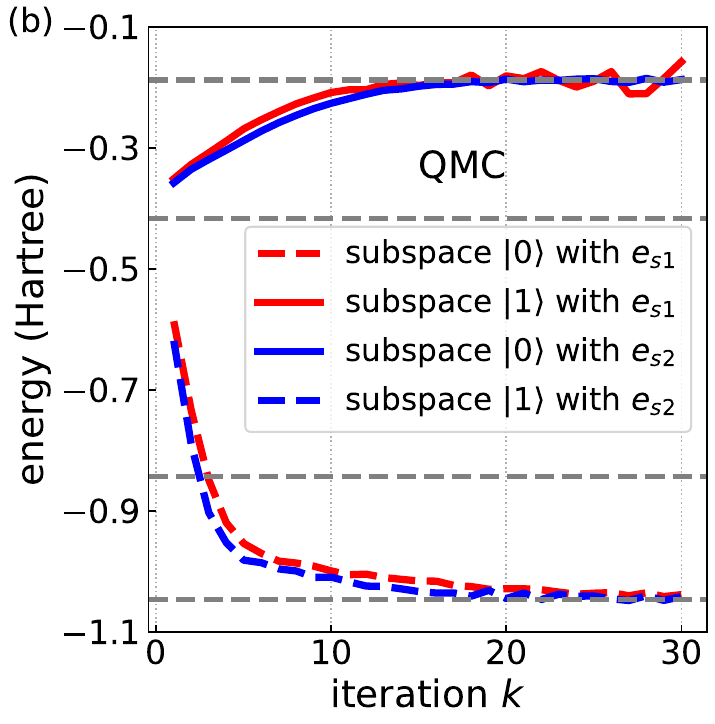}
  \caption{Simulation of the hydrogen molecule at an inter-nuclear distance $R=1.25$ Angstrom in the LCU (a) and QMC (b) methods. The energy-shift $e_s$ is set to a value near different eigenstates such that the measurement outputs converge to the corresponding eigenenergy when increasing iterations. The gray horizontal dashed lines are the theoretical spectra while the solid curves are the simulation outputs.}
    \label{fig2_H2} 
 \end{figure}

\begin{figure*}[ht]
    \centering
\includegraphics[width=\linewidth]{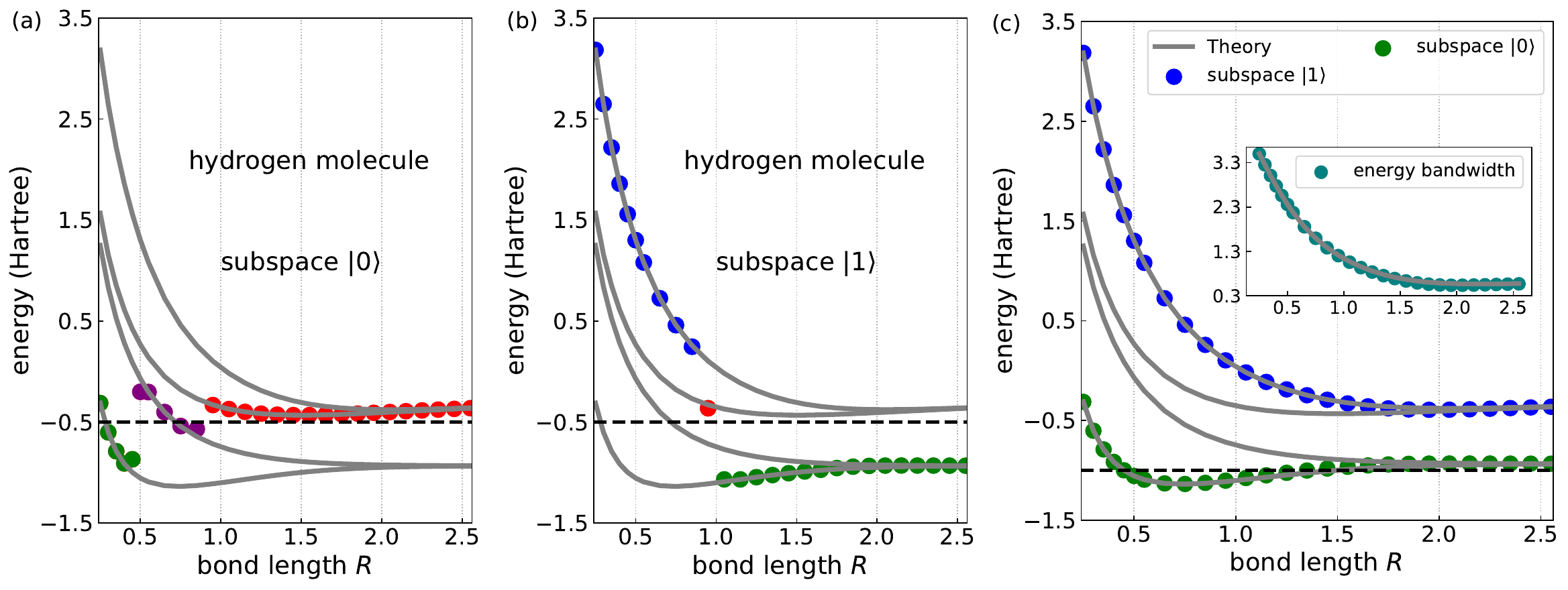}
  \caption{Performance of our quantum algorithm in different subspaces when varying the inter-nuclear distance of a hydrogen molecule with a fixed energy-shift $e_{s}=-0.5$ (a,b) and $e_s=-1$ (c). The gray  curves are the theoretical spectra under different inter-nuclear distances, while the circular points denote the simulation outputs. The energy bandwidth is obtained via the difference between the highest excited state and the ground state.}
    \label{fig3_attractor} 
 \end{figure*}
 
The fermionic Hamiltonian of the hydrogen molecule can be translated into qubit representation by the Bravyi-Kitaev transformation~\cite{bravyi2002fermionic} and its two-qubit Hamiltonian is 
\begin{equation}
\begin{split}
H(R)=&\alpha_{0}^{R}+\alpha_{1}^{R}\sigma_{z}^{(1)}+\alpha_{2}^{R}\sigma_{z}^{(2)}+\alpha_{3}^{R}\sigma_{z}^{(1)}\otimes\sigma_{z}^{(2)} \\
&+\alpha_{4}^{R}\sigma_{x}^{(1)}\otimes\sigma_{x}^{(2)}+\alpha_{5}^{R}\sigma_{y}^{(1)}\otimes\sigma_{y}^{(2)},
\end{split}
\end{equation}
where $\sigma_{\beta}^{(i)}~(\beta=x,y,z)$ is the Pauli operator acting on the $i$-th qubit and the real-valued coefficients $\alpha_{i}^{R}$ are functions of the inter-nuclear distance~\cite{colless2018computation}. 
    
We first demonstrate the algorithm using the LCU method at the inter-nuclear distance $R=1.25$ Angstrom, with four eigenenergies $\{-1.0458,-0.8428,-0.4166,-0.1878\}$ in Fig.\ref{fig2_H2}\textcolor{blue}{(a)}. The initial state is chosen as $(\ket{00}+2\ket{01}+\ket{10}+\ket{11})/\sqrt{7}$, with initial values of the energy-shift parameters $e_{s1}=-1.1$ and $e_{s2}=0$, and the evolution time is set as $t=1.3518$. We can see that with increasing iterations, the output energies rapidly evolve towards the nearest (or farthest) eigenenergy relative to the energy-shift in subspace $\ket{0}$ (or subspace $\ket{1}$). The theoretical expectation of the energy bandwidth is $0.8580$, while the algorithm output is $0.8565$ with error $1.5\times10^{-3}$, reaching chemical accuracy. 

Moreover, we provide the simulation results in the QMC method with same parameter setup, as shown in Fig.\ref{fig2_H2}\textcolor{blue}{(b)}. We choose the observation quantity $O=H$ for comparison and the relevant parameters are set as error $\epsilon=0.05$ and the number of samples is $n_{\rm{s}}=2^{24}$. We found that this method can also indeed work, and once an $e_s$ is selected within an energy interval, as the number of iterations $k$ increases, the system will gradually approach the target energy eigenvalues in different subspaces and the algorithm output of energy bandwidth is $0.8530$. Furthermore, we fix the value of the energy-shift $e_s=-0.5$ and vary the inter-nuclear distance $R$. From Fig.\ref{fig3_attractor}\textcolor{blue}{(a,b)}, we observe that the final simulation results (iteration = 30) at each point converge to the eigenenergy level closest to or farthest from the energy shift, acting as either an attractor in the  $\ket{0}$ subspace, or as a repeller in the $\ket{1}$ subspace. As shown in Fig.\ref{fig3_attractor}\textcolor{blue}{(c)}, if we set $e_s=-1$, we can obtain the energy bandwidth between the ground state and the highest exited state.

\subsection{Analysis of the algorithm features}

In this section, we focus on the subspace $\ket{0}$ of the quantum system where the TSQES algorithm reduces to the QESDS algorithm. In the traditional quantum methods for solving excited states~\cite{jones2019variational,higgott2019variational,wen2024full,wang2024improving}, the target state is approximated by solving the lower energy levels step-by-step, and updating the Hamiltonian, which will transfer the error of the lower energy space to the excited state, resulting in error-accumulation. However, the mechanism of our QESDS algorithm is to directly solve specific energy levels and this can effectively avoid this problem. 

\subsubsection{Avoiding the error-accumulation phenomenon}

\begin{figure}
    \centering
\includegraphics[width=0.85\linewidth]{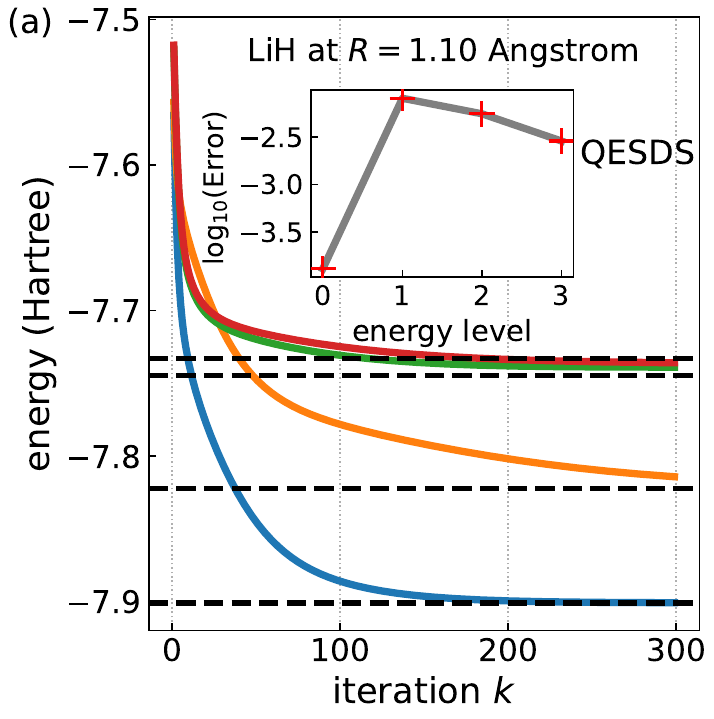}
\includegraphics[width=0.85\linewidth]{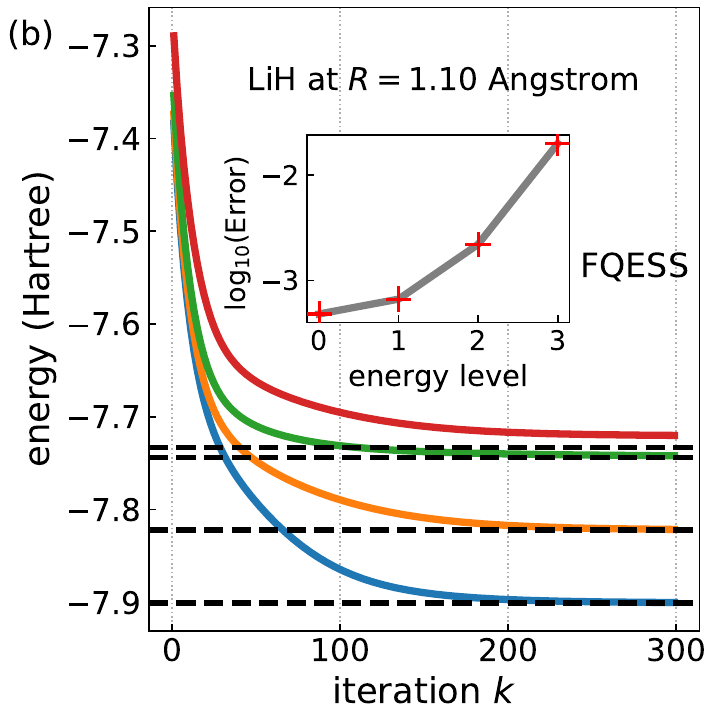}
  \caption{Simulation of the LiH molecule at an inter-nuclear distance $R=1.10$ Angstrom. The black dashed horizontal lines indicate the target energy levels obtained through matrix diagonalization, while the solid curves depict the numerical simulation results of the QESDS  (a) and FQESS (b) algorithms.}
    \label{fig4_error} 
 \end{figure}  
 
We now apply the QESDS and FQESS algorithms~\cite{wen2024full} to a six-qubit LiH molecule at an inter-nuclear separation of 1.10 Angstrom, as shown in Fig.\ref{fig4_error}. The initial state is chosen as $|+ \rangle ^{\otimes 6}$, where $|+ \rangle$ is an eigenvector of the $\sigma_x$ Pauli matrix, and the three lowest eigenenergies are calculated. We can clearly find that the error between the output results of the FQESS algorithm and the theoretical expectations gradually increases when increasing the energy levels. However, similar error-accumulation effects do not occur in the QESDS algorithm, due to the independence of the solving process for different energy levels. 

\subsubsection{Impact of the evolution time}

Furthermore, we now investigate the relationship between the convergence rate and the evolution time $t$. In fact, the superposition of the forward and backward time evolution in the QESDS algorithm forms a cosine filtering operator, whose peak is determined by the energy-shift $e_s$. Mathematically, the absolute value of the gradient for the cosine function $\cos(x)$ with $x \in [-\pi/2, \pi/2]$ decreases first and then increases with a minimum at $x=0$, around which the corresponding change in the dependent variables is minimal. Moreover, the change in $(E_{i}-e_{s})t$ is proportional to $|E_{i}-e_{s}|$ as $|t|$ increases. 

As shown in Fig.\ref{fig5_cosine}\textcolor{blue}{(a)}, the increase of $|t|$ leads to a dual amplification effect on the amplitude difference of the quantum state components. In other words, if the energy-shift $e_s$ is closest to the eigenenergy $E_m$, then the differences $ \cos((E_{m}-e_{s})t)-\cos((E_{m-1}-e_{s})t)$ and $\cos((E_{m}-e_{s})t)- \cos((E_{m+1}-e_{s})t)$ will both increase due to the increase in $| t | $, and consequently accelerating the convergence rate. Therefore, we can naturally point out that within the constraints of $(E_{i}-e_{s})t \in [-\pi/2,\pi/2]~\forall~i$, the QESDS algorithm converges faster when $t$ approaches a critical value. Actually, this aligns with intuition, because we can regard $t$ as the training step size, which is one of the main factors affecting the iteration speed. We take the calculation of the ground state and the fifth excited state of the LiH molecule as an example, as shown in the Fig.\ref{fig5_cosine}\textcolor{blue}{(b)}. We can conclude that the convergence becomes faster and fewer iterations are required as the evolution time increases. 

\begin{figure}
        \centering
        \includegraphics[width=\linewidth]{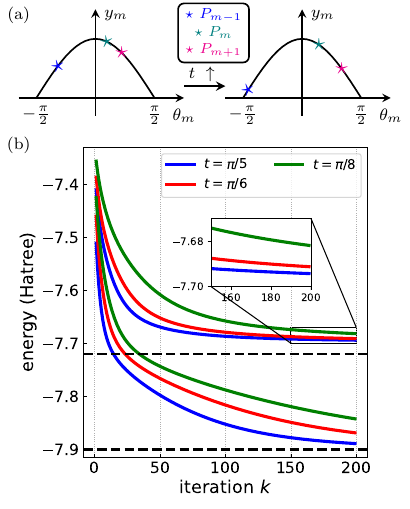}
        \caption{(a) Schematic diagram of the cosine function illustrating the impact of the evolution time on the convergence rate. We here set $\theta_m=(E_m-e_s)t$ and $y_m=\cos(\theta_m)$. The point $P_m=(\theta_m,y_m)$ is labeled by stars. (b) Numerical simulations of the QESDS algorithm under different evolution times. The target eigenstates are the ground state with eigenvalues $-7.9008$ and $e_s = -8.0$, and the fifth excited state with eigenvalues $-7.7202$ and $e_s = -7.7$. The black dashed horizontal lines indicate the target energy levels, while the solid curves depict the simulation results.}
        \label{fig5_cosine}
        \end{figure}

\begin{figure}
        \centering
        \includegraphics[width=\linewidth]{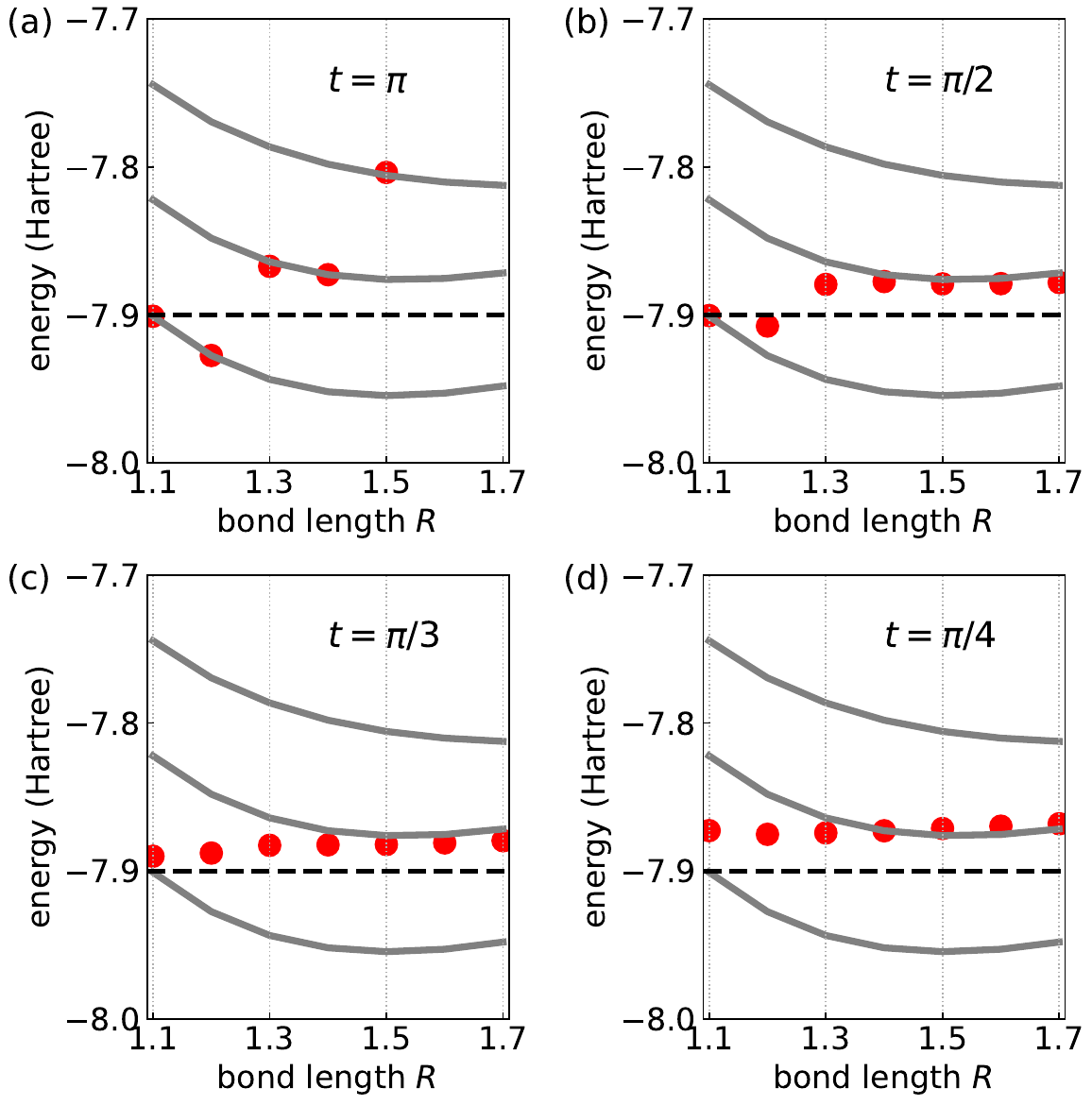}
        \caption{Energy spectra of the LiH molecule with fixed energy-shift $e_s$ for different evolution times. The gray curves represent the energy levels for the ground, first excited, and second excited states, obtained from classical diagonalization; while the dashed black horizontal line represents the value of the energy-shift. The red circular points denote the simulation results obtained using the QESDS algorithm. }
        \label{fig6_LiHt}
        \end{figure}
        
In addition, we attempt to relax the restriction of $(E_{i}-e_{s})t \in [-\pi/2,\pi/2]~\forall~i$ to further accelerate the convergence rate. Figure \ref{fig6_LiHt} illustrates the results of our algorithm for a series of nuclear distances in the LiH molecule with $e_s = -7.9$ fixed and $k=400$ under different evolution time parameters. Though the evolution time $t \in \{\pi, \pi/2, \pi/3, \pi/4\}$ does not meet the aforementioned constraint, the optimal performance is still achieved at $t=\pi/2$, where the first two simulation points fall onto the ground state and the subsequent points correspond to the first excited state energy. However, the algorithm may not work as expected if $t$ is too large. For example, the last two points fall onto eigenvalues far from $e_s$ when $t=\pi$. This demonstrates that sometimes it is appropriate to relax the constraints on evolution time for faster convergence, but the success of the constraint relaxation is based on the condition that the amplification of all the other components is still smaller than that of the target eigenstate, i.e., $|\cos(E_{i}-e_{s})t|$ takes its maximum value when $E_i$ is closest to $e_s$.


\section{Application of the quantum excited state direct solver to condensed matter systems}

In this section, we apply our QESDS algorithm to condensed matter systems and investigate the interesting phenomenon of topological phases. Generally, there exists a significant energy gap known as forbidden band between the conduction and valence bands in the energy band structure of insulators, and the Fermi level resides within this band gap for insulators. Topologically non-trivial insulators can be regarded as a new system bridging trivial insulators and low-dimensional metals~\cite{qi2011topological}, whose feature edge states are protected by topology within the band gap, a fundamentally different property from trivial insulators. Therefore, the band structure at the energy gap of insulators distinguishes trivial insulators from topologically non-trivial insulators. 

\begin{figure}
    \centering
        \includegraphics[width=0.9\linewidth]{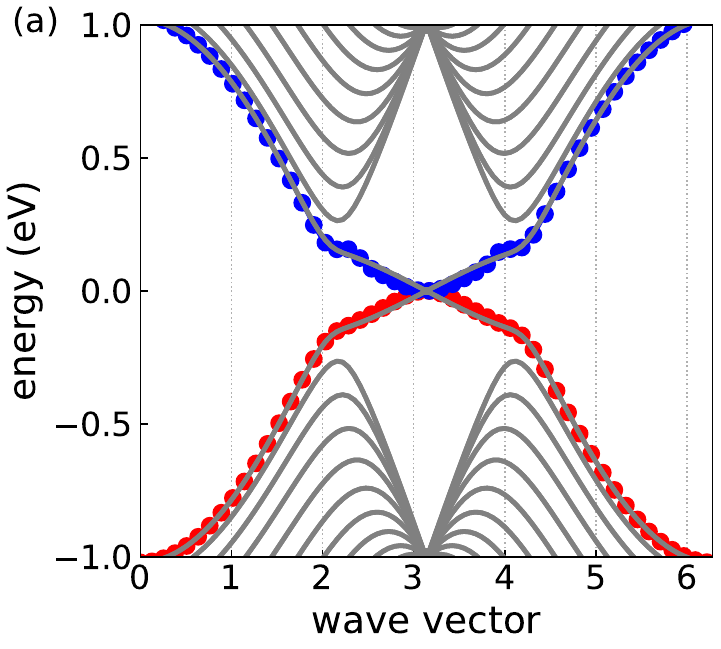}
            \includegraphics[width=0.9\linewidth]{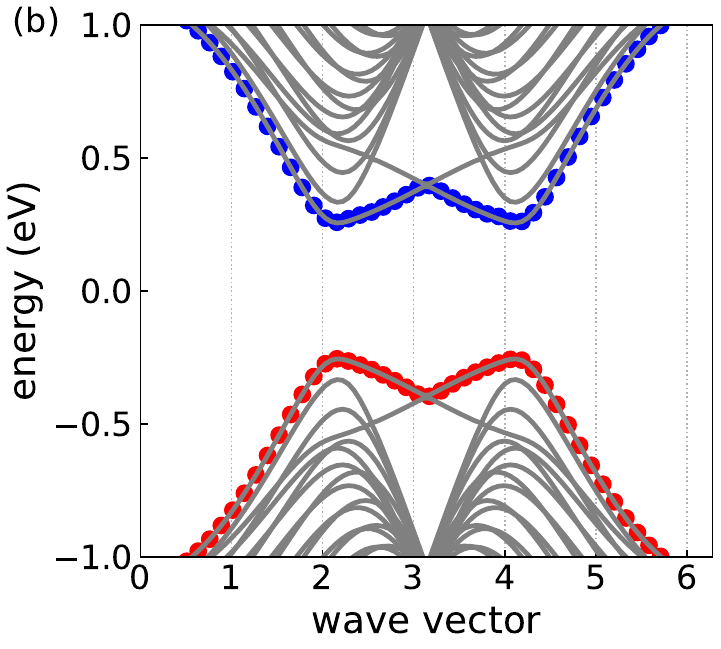}
    \caption{Simulation results of the Kane-Mele model in the topological edge states (a) and in the trivial states (b). The gray solid curves represent the band structures by classical computations, while the colored circular points represent the simulation results of the QESDS algorithm.}
    \label{fig7_KM}
\end{figure}


\subsection{Distinguishing quantum phases of the $Z_2$ topological insulator}

An important theoretical model for realizing a two-dimensional $Z_2$ topological insulator is the Kane-Mele model~\cite{kane2005z,kane2005quantum}. It arises from considering spins as a variable in the graphene lattice and can be regarded as a combination of two independent Haldane Hamiltonians, each corresponding to spins pointing up and down~\cite{haldane1988model}. The Hamiltonian can be written as~\cite{kane2005quantum}
\begin{equation}\label{equ:kanehamiltonian}
\begin{split}
            H_{km}&= t_1 \sum_{\langle i,j \rangle s}c_{i,s}^{\dagger}c_{j,s} + i t_2\sum_{\langle \langle i,j \rangle\rangle ss'} v_{ij} s_z c_{i,s}^{\dagger}c_{j,s'} \\ 
             + &i t_3\sum_{\langle \langle i,j \rangle\rangle ss'} c_{i,s}^{\dagger}\left(\vec{s} \times \vec{d} \right)_z c_{j,s'}  + M \sum_i c_{i,s}^{\dagger}c_{i,s},
\end{split}
\end{equation}
where the first term represents electron hopping between nearest-neighbor lattice sites, the second term represents the electron spin-orbit coupling between next-nearest neighbor lattice sites, the third term is a nearest-neighbor Rashba term, which explicitly violates the $z \rightarrow -z$ mirror symmetry, and the last term is the staggered mass term. The subscripts $s$ and $s'$ denote spin-up and spin-down, respectively. The $v_{ij}$ indicates the direction of the next-nearest neighbor transition: $v_{ij} = +1$ for a clockwise transition and  $v_{ij} = -1$ for a counterclockwise transition. 

Neglecting the Rashba spin-orbit coupling term, we take a Kane-Mele model consisting of 20 unit cells  as an example. The dimension of the Hamiltonian is $160$, and two values at the energy gap are selected as the energy-shift $e_s\in\{-0.1,0.1\}$. The evolution time parameter is set as $t=\pi/5$ and the number of iterations is $k=150$. The simulation results are shown in Fig.\ref{fig7_KM}, demonstrating that our algorithm aligns well with theoretical expectations. We can find that for $t_1 = 1$, $t_2 =0.03$, $M=0$, two intersecting lines exist within the energy gap, indicating topologically protected edge states. However, when the mass parameter is varied to $M=0.4$, these topological edge states vanish, resulting in a band structure characteristic of a trivial non-topological insulator.



\subsection{Distinguishing quantum phases of Su-Schrieffer-Heeger model}

\begin{figure*}
    \centering
    \includegraphics[width=\linewidth]{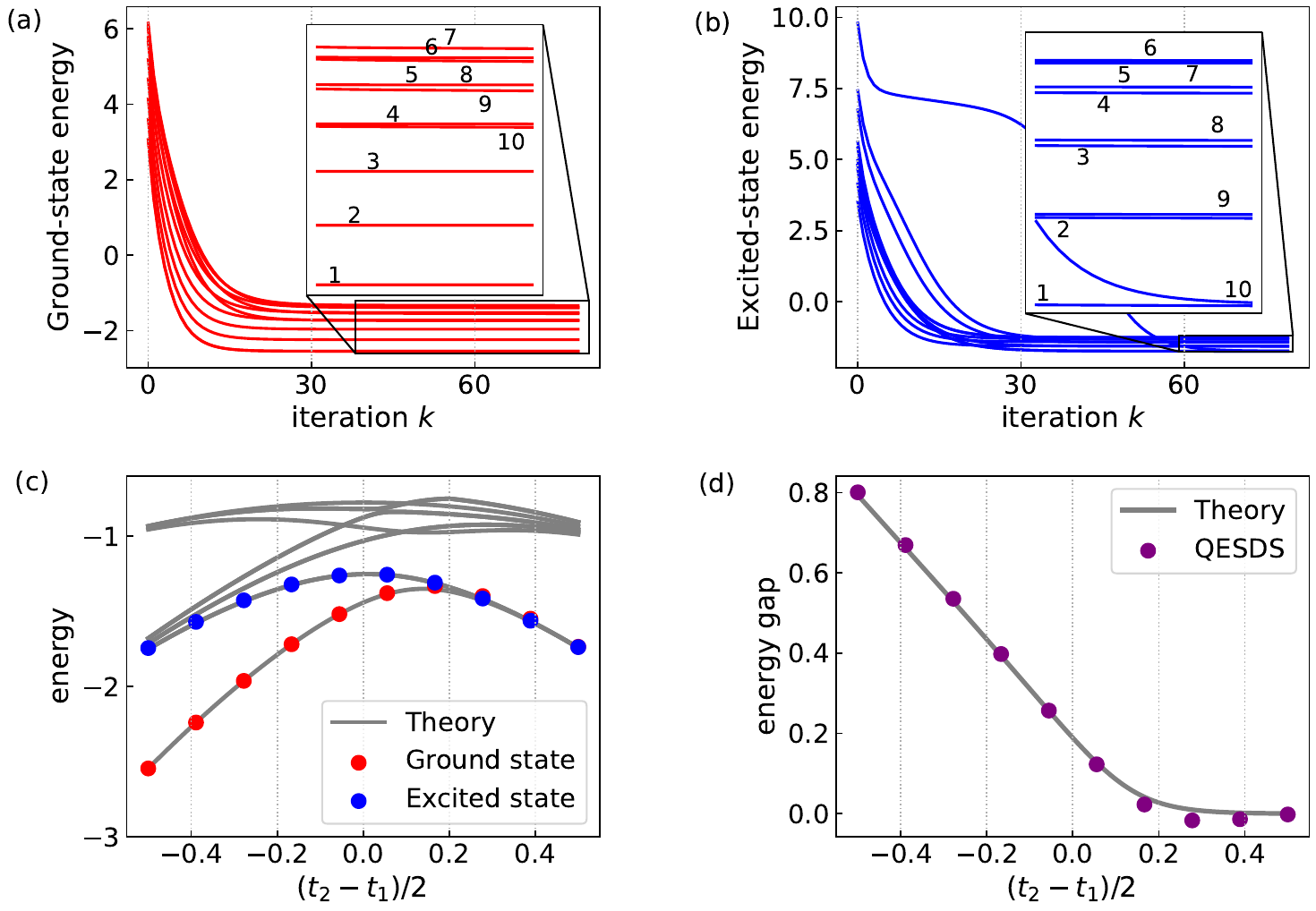}
    \caption{Simulation results of the eigenenergies of the SSH model with Hubbard interaction at half-filling ($U=10,\ N=6$) with our QESDS algorithm. (a-b) Energy evolution versus with the iteration number $k$ of the ground state and the first excited state. The numbers 1 to 10 represent the ten curves with increasing $(t_2-t_1)/2$ values. (c) The gray curves represent the band structures by classical exact diagonalization, while the circular color dots denote the results obtained by our QESDS algorithm. The initial states and energy-shift parameters $e_s$ are chosen as the eigenstates and eigenenergies of the SSH model without Hubbard interaction ($U=0$). (d) The  energy gap ($E_1-E_0$) exhibits a dependence on the difference between the inter-cell and intra-cell hopping. }
    \label{fig8_SSH}
    \end{figure*}
    
Moreover, determining the topological state of a system under strong electronic correlations is also a fundamental issue in the study of topological physics. Here, we now introduce a strong correlation term, the Hubbard interaction, into the Su-Schrieffer-Heeger (SSH) model. It is well-established that the SSH model exhibits a topological state when the inter-cell hopping exceeds the intra-cell hopping. However, the inclusion of Hubbard interaction can significantly influence the topological properties of the ground state. To address this, we focus on the system at half-filling and analyze the edge states under a large interaction strength to infer its topological characteristics. The Hamiltonian of the SSH with Hubbard interaction is~\cite{jiang2012identifying,le2020topological}
\begin{equation}
    \begin{aligned}
        H= &  -\sum_{i,\sigma}(t_1\hat{c}_{i,\sigma}^{A\dagger}\hat
        {c}_{i,\sigma}^{B}+t_2\hat{c}_{i,\sigma}^{B\dagger}\hat{c}_{i+1,\sigma}
        ^{A}+h.c.)+\sum_{i,s} U\hat{n}^{s}_{i\uparrow}\hat{n}^{s}_{i\downarrow},
        \newline
    \end{aligned}
\end{equation}
where $\hat{c}_{i,\sigma}^{s\dagger}$ is the fermion creation operator of the $i$-th site with sub-lattice label $s$ and spin label $\sigma$. The parameters $t_2$ and $t_1$ represent the nearest-neighbor inter-cell and intra-cell hopping, respectively, and $U$ is the on-site interaction strength. For $\hat{d}_{i,1}=\hat{c}_{i\uparrow}^A$, $\hat{d}_{i,2}=\hat{c}_{i\uparrow}^B$, $\hat{d}_{i,3}=\hat{c}_{i\downarrow}^A$, and $\hat{d}_{i,4}=\hat{c}_{i\downarrow}^B$, we have
\begin{equation}
    \begin{aligned}
        H= &  -\sum_{i}[
        t_1(\hat{d}_{i,1}^{\dagger}\hat{d}_{i,2}+\hat{d}_{i,3}^{\dagger}\hat{d}_{i,4})+
        t_2(\hat{d}_{i,2}^{\dagger}\hat{d}_{i+1,1}+\hat{d}_{i,4}^{\dagger}\hat{d}_{i+1,3})
        \\
        &+h.c.] +\sum_{i} U(\hat{d}_{i, 1}^{\dagger}\hat{d}_{i, 1}\hat{d}_{i, 3}^{\dagger}\hat{d}_{i, 3}+\hat{d}_{i, 2}^{\dagger}\hat{d}_{i, 2}\hat{d}_{i, 4}^{\dagger}\hat{d}_{i, 4}),\newline
    \end{aligned}
\end{equation}

We construct a chain with $N=6$ sites and with six electrons and focus on the properties of the ground state and excited states. We take the computable eigenstates and eigenvalues of the Hamiltonian without Hubbard interaction ($U=0$) as the initial state and the energy-shift values $e_s$ of the QDEDS algorithm, so that the algorithm results in the presence of strong many-body interactions can be well calculated. The evolution of low-energy states, calculated by the classical exact diagonalization method and the QESDS algorithm, with respect to $\Delta_t=(t_2-t_1)/2$ at a hopping difference $U=10$ is shown in Fig.\ref{fig8_SSH}.

We can conclude that the simulation results agree well with the theoretical expectations. As the difference between the inter-cell hopping ($t_2$) and intra-cell hopping ($t_1$) increases, the degeneracy of the lowest energy (ground) state transitions from one to four. And we can further obtain the degeneracy of the ground state by calculating the energy gap~\cite{le2020topological}, defined as the gap between the first exited state energy and the ground state energy $E_{\mathrm{gap}} = E_1- E_0$. It can be founded that there is a topological phase transition occurs when the energy gap reaches zero, which indicates the presence of edge states and the system is in a topologically non-trivial phase.

Moreover, we emphasize that our QESDS algorithm can provide a distinctive advantage by enabling direct computation of the band structure at the energy gap without concerning other bands. This feature renders it convenient for assessing the topological properties of insulators compared to other quantum algorithms. Additionally, by setting an energy-shift $e_s$, the presence and width of the band gap can be directly determined, allowing for differentiation between the band structures of conductors, semiconductors, and insulators. We offer a demonstration to this point with the Bistritizer-MacDonald model~\cite{bistritzer2011moire} in the Appendix \ref{appenBMmodel}.


\section{Conclusion} \label{Chapconclusion}

In this work, we use the time-symmetry property of quantum mechanics to design a quantum eigensolver. Utilizing the superposition of forward and backward time evolution operators with an energy-shift parameter, we realize the TSQES algorithm
and its reduced version, QESDS, which enable us to directly solve any eigenstate in quantum many-body systems without the need for prior determination of lower eigenstates or eigensubspaces. 

The error-accumulation effect in the process of solving high-energy excited states can be avoided, and the algorithm can converge to the target states quickly and robustly because of its non-variational character. The feasibility of realizing such non-unitary evolution processes is presented with the LCU and QMC methods, respectively. And we also improve the algorithm to a non-iterative full quantum circuit form by adding ancillary qubits, which reduces the resource cost caused by repeated quantum state measurements.

We have successfully applied the TSQES algorithm to calculate the eigenenergies of quantum chemical systems, including hydrogen and LiH molecules. Moreover, we show the significant advantages of the QESDS algorithm brought by the mechanism of directly finding specific eigenstates in the determination of topological states in condensed matter systems, including the Kane-Mele model and half-filled SSH with Hubbard interaction. Our algorithm presents a novel technical framework for studying high-energy excited states and specific energy band structures using quantum computers, offering new possibilities for quantum simulations of many-body quantum systems.

The performance of our algorithm demonstrates that forward and backward time evolution offers a computational advantage over methods that rely on a fixed time direction. Time symmetry provides a more accurate description of the underlying physics.
The design framework of quantum algorithms based on time symmetry can be extended to various fields. By expanding the Hilbert space, we can continuously control the constructive and destructive interference between forward and backward time evolution, enabling the realization of more non-unitary, complex evolutionary processes, including those in open systems.


\section{Acknowledgements}
S.W. acknowledges the Beijing Nova Program under Grants No. 20230484345 and 20240484609; F.N. is supported in part by the Japan Science and Technology Agency (JST) via the CREST Quantum Frontiers program Grant No. JPMJCR24I2, the Quantum Leap Flagship Program (Q-LEAP), and the Moonshot R\&D Grant Number JPMJMS2061; We also acknowledges the National Natural Science Foundation of China under grant
No. 62471046.


\appendix

\section{Arbitrary observable estimation\label{appendix_A}}

\subsection{Estimating the Denominator $D$}

At first, we will discuss estimating $D$. We define 
\begin{align}
\overline{\hat{D}}&=\frac{1}{n_D}\sum_{i=1}^{n_D}\hat{D},\\
\overline{\hat{N}}&=\frac{1}{n_N}\sum_{i=1}^{n_N}\hat{N},
\end{align}
so it is obvious that $\mathbb{E}{\overline{\hat{D}}}=D,\mathbb{E}{\overline{\hat{N}}}=N$.

For the estimation precision $\epsilon_D$ of the denominator $D$, usually its estimation needs to be divided into real part $\Re[\cdot]$ and imaginary part $\Im[\cdot]$ separately, and we assume that they are selected with equal probability, then according to the Hoeffding inequality~\cite{zeng2021universal},
\begin{align}
\mathrm{Pr}(|\Re\overline{\hat{D}}-\mathbb{E}\Re\overline{\hat{D}}|\geqslant\epsilon_D)\leqslant 2\exp{\left(-\frac{n_D\epsilon_D^2}{2}\right)},
\end{align}
and the imaginary part $\Im[\cdot]$ can be estimated in the same way. So the sample number of the denominator will be
\begin{align}\label{n_D}
n_D\geqslant \frac{2\cdot K_D}{\epsilon_D^2},
\end{align}
with a total failure probability of the denominator estimation
\begin{align}
\delta_D\leqslant 4\exp{\left(-\frac{K_D}{2}\right)},
\end{align}
where $K_D>0$ is a bound parameter for manually controlling the sampling number and failure probability. 

\begin{figure}
  \centering
  \includegraphics[width=\linewidth]{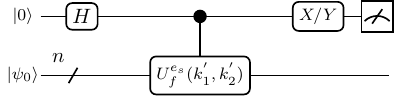}
  \caption{Quantum circuit for measuring the denominator estimator $\hat{D}_{k_1^{'},k_2^{'}}$.}
  \label{fig9_circuitD}
\end{figure}

The circuit diagram for measuring the denominator estimator $\hat{D}_{k_1^{'},k_2^{'}}$ is shown in Fig.\ref{fig9_circuitD}. There, $U_f^{e_s}(k_1^{'},k_2^{'})=(U_f^{e_s})^{k_1^{'}-k_2^{'}}$, where $k_1^{'},k_2^{'}$ are generated separately by the binomial distribution $B(k,1/2)$. When the initial state $|\psi_0\rangle$ passes through this quantum circuit, the state will be
\begin{align}
|\Psi_n^D\rangle=\frac{1}{\sqrt{2}}(|0\rangle\otimes|\psi_0\rangle+|1\rangle\otimes(U_f^{e_s})^{k_1^{'}-k_2^{'}}|\psi_0\rangle).
\end{align}
Therefore, 
\begin{align}
\langle\Psi_n^D|X|\Psi_n^D\rangle&=\mathrm{Re}\left(\langle\psi_0|(U_f^{e_s})^{k_1^{'}-k_2^{'}}|\psi_0\rangle \right), \nonumber\\
\langle\Psi_n^D|Y|\Psi_n^D\rangle&=\mathrm{Im}\left(\langle\psi_0|(U_f^{e_s})^{k_1^{'}-k_2^{'}}|\psi_0\rangle \right). 
\end{align}
After that, the estimator $\hat{D}(k_1^{'},k_2^{'})$ can be obtained by calculating the expectation about $(k_1^{'},k_2^{'})$:
\begin{align}\label{Dk1k2}
\hat{D}(k_1^{'},k_2^{'})=\left(\langle\Psi_n|X|\Psi_n^N\rangle+i\langle\Psi_n|Y|\Psi_n^N\rangle\right),
\end{align}
and then 
\begin{align}\label{D}
D=\mathbb{E}_{k_1^{'},k_2^{'}}\hat{D}(k_1^{'},k_2^{'}).
\end{align}
Finally, the algorithm for estimating $D$ is given in Algorithm \ref{alg:D}.%

\begin{algorithm}[H]
\caption{Estimating the denominator $D$}
\label{alg:D}
\begin{algorithmic}[1]

\vspace{0.2cm}
\Require  $U_f^{e_s}=\exp\left({-i(H-e_s)t}\right)$; initial state $|\psi_0\rangle$ with nonzero overlap with $j$-th eigenstate of Hamiltonian $H$: $|a_j|^2 = |\langle E_j | \psi_0 \rangle|^2\neq0$; the interval $\left[E_j^L, E_j^U\right]$ of $E_j$, and the energy-shift $e_s$ selected in $\left[E_j^L, E_j^U\right]$; 

\noindent estimation accuracy $\epsilon$ of the Hamiltonian eigenstates $E_j$, the total failure probability bound $\delta$, and selecting the iteration times 
\[
k=\log\left(\frac{\|H\|_1}{\epsilon|a_s|^2}\right)/{2\log\left(\frac{\lambda_s}{\lambda_t}\right)}
\]
according to Eq.(\ref{kk}); the properly selected sampling numbers 
\[
n_D=O\left(\frac{\ln(\frac{1}{\delta})\cdot(\|O\|_1+1)^2}{|a_s|^4\cdot\epsilon^{-4c\cdot \log\left(\lambda_s\right)}}\right).
\]
\Ensure  Estimation of the numerator $D$ with failure probability $\delta_D\leqslant \delta/2$.
\For {$q = 1$ to $n_D$}
    \State Sample two non-negative integers $k_1^{'}, k_2^{'}$ separately according to the binomial distribution $B(k,1/2)$.
    \State Running the quantum circuit diagram of the Fig.\ref{fig9_circuitD} to obtain the final state $|\Psi_n^D\rangle$.
    \State Implement the measurement $M$ to obtain $\langle\Psi_n^D|M|\Psi_n^D\rangle$, $M\in\{X,Y\}$, and  is selected sequentially.
\EndFor
    \State Calculate $\hat{D}(k_1^{'},k_2^{'})$ according to Eq.(\ref{Dk1k2}).
\State Calculate the estimated denominator $\overline{\hat{D}}$ as an estimation of $D$: $\overline{\hat{D}}= \frac{1}{n_D} \sum_{p=1}^{n_D} \hat{D}(k_1^{'},k_2^{'})$.
\end{algorithmic}
\end{algorithm}

\subsection{Estimating the Numerator $N(O)$}

For the estimation precision $\epsilon_N$ of the numerator $N(O)$, in a similar way as the estimation of $D$, according to the Hoeffding inequality, we can obtain that
\begin{align}
\mathrm{Pr}(|\overline{\Re\hat{N}}-\mathbb{E}\Re\overline{\hat{N}}|\geqslant\epsilon_N)\leqslant 2\exp{\left(-\frac{n_N\epsilon_N^2}{2\|O\|_1^2}\right)},
\end{align}
so the sample number of the denominator will be
\begin{align}\label{n_N}
n_N\geqslant \frac{2\cdot K_N\cdot\|O\|_1^2}{\epsilon_N^2},
\end{align}
with a failure probability 
\begin{align}
\delta_N\leqslant 4\exp{\left(-\frac{K_N}{2}\right)},
\end{align}
where $K_N>0$ is also a bound parameter for manually controlling the sampling number and failure probability.

Finally, taking into account the failure probabilities of the numerator $N(O)$ estimation and denominator $D$ estimation, the total failure probability for estimating $\langle O\rangle=N(O)/D(O)$ becomes 
\begin{align}
\delta\leqslant\delta_D+\delta_N=8\exp{\left(-\frac{K}{2}\right)}
\Leftrightarrow K\leqslant 2\ln\left(\frac{8}{\delta}\right),
\end{align}
where we have assumed that $K_D=K_N=K$ for the sake of convenience.

\subsection{Sampling complexity analysis}

Based on the analysis of the sampling complexity of the numerator and denominator, the total error estimation of the observed quantity can be performed as follows:
\begin{align}
\left|\langle O\rangle-\langle\hat{O}\rangle\right| & =\left|\frac{N(O)}{D}-\frac{\overline{\hat{N}}(O)}{\overline{\hat{D}}}\right| \nonumber\\
& =\left|\frac{N(O) \overline{\hat{D}}-D \overline{\hat{N}}(O)}{D \hat{D}}\right| \nonumber\\
& =\left|\frac{N(O) (\overline{\hat{D}}-D)+D \left(\overline{\hat{N}}(O)-N(O)\right)}{D\left(D+(\overline{\hat{D}}-D)\right)}\right| \nonumber\\
& =\left|\frac{\langle O\rangle (\overline{\hat{D}}-D)+ \left(\overline{\hat{N}}(O)-N(O)\right)}{D+(\overline{\hat{D}}-D)}\right| \nonumber\\
& \leqslant\frac{|\langle O\rangle|\epsilon_D+\epsilon_N }{D-\epsilon_D} \nonumber\\
&\leqslant\frac{(|\langle O\rangle|+1)\epsilon_D+\epsilon_N }{D} \nonumber\\
&\leqslant\frac{(\|O\|_1+1)\epsilon_D+\epsilon_N }{D} \nonumber\\
&\leqslant\frac{(\|O\|_1+1)\epsilon_D+\epsilon_N }{|a_s|^2\lambda_s^{2k}},
\end{align}
where we have made the reasonable assumption that $N(O)\epsilon_D$ and $D\epsilon_N$ meet the following condition 
\begin{align}
&(|\langle O\rangle|+1)\epsilon_D+\epsilon_N\ll D \nonumber\\
\Longrightarrow& \epsilon_D\ll D/(|\langle O\rangle|+1), \epsilon_N\ll D.
\end{align}
Since $|a_s|^2\lambda_s^{2k}\ll D$ always holds, we can replace the above equation with the following more stringent condition
 \begin{align}
&(\|O\|_1+1)\epsilon_D+\epsilon_N\ll |a_s|^2\lambda_s^{2k}   \nonumber\\
\Longrightarrow &\epsilon_D\ll |a_s|^2\lambda_s^{2k}/(\|O\|_1+1), \epsilon_N\ll |a_s|^2\lambda_s^{2k}.
\end{align}
then according to the Eq.(\ref{n_D}), Eq.(\ref{n_N}) we know that $n_D$, $n_N$ should meet the following condition 
\begin{align}
n_D&\geqslant \frac{2K_D}{\epsilon_D^2}\gg\frac{2K_D\cdot(\|O\|_1+1)^2}{|a_s|^4 \lambda_s^{4k}}, \\
n_N&\geqslant \frac{2K_N\cdot\|O\|_1^2}{\epsilon_N^2}\gg \frac{2K_N\cdot\|O\|_1^2}{|a_s|^4 \lambda_s^{4k}}.
\end{align}
Therefore, in order to meet the above conditions, recall that $\lambda_s\in(1/2^q,1)$, $q\in(0,1)$, the sample complexity of the denominator $D$ and numerator $N(O)$ must be
\begin{align}
n_D&=O\left(\frac{K_D\cdot(\|O\|_1+1)^2}{|a_s|^4 \lambda_s^{4k}}\right)=O\left(\frac{K_D\cdot(\|O\|_1+1)^2}{|a_s|^4\cdot\epsilon^{-4c\cdot \log(\lambda_s)}}\right)\nonumber\\
&=O\left(\frac{\ln(\frac{1}{\delta})\cdot(\|O\|_1+1)^2}{|a_s|^4\cdot\epsilon^{4c\cdot q}}\right),
\label{n_D_order}\\
n_N&=O\left(\frac{K_N\cdot\|O\|_1^2}{|a_s|^4 \lambda_s^{4k}}\right)=O\left(\frac{K_N\cdot\|O\|_1^2}{|a_s|^4\cdot\epsilon^{-4c\cdot \log(\lambda_s)}}\right)\nonumber\\
&=O\left(\frac{\ln(\frac{1}{\delta})\cdot\|O\|_1^2}{|a_s|^4\cdot\epsilon^{4c\cdot q}}\right).
\label{n_N_order}
\end{align}
From the above formulas, we can clearly see the dependence of the sampling complexity of $n_D$, $n_N$ on the number of iterations $k$, the total failure probability bound, and truncation accuracy $\epsilon$, which provides us with guidance for selecting $\epsilon$, $k$, and even $\lambda_s$ ($e_s$). Note that in actual sampling, the number of samples is often much smaller than the number specified in the above formula to achieve a considerable level of accuracy.

Next, let us take a look at the quantum circuit diagram for implementing $N(O)$. 

The circuit diagram for measuring the numerator estimator $\hat{N}_{n,k_1,k_2}$ is shown in Fig.\ref{fig10_circuitN}. This circuit is actually a Hadamard test circuit. There, $U_f^{e_s}(k_x)=(U_f^{e_s})^{2k_x-k}$, where $x\in\{0,1\}$, and $k_x$ is generated by the binomial distribution $B(k,1/2)$. 

In order to realize the quantum circuit directly, the observable is divided into a linear superposition of a series of generalized Pauli operators $\sigma_n$, i.e., $O=\sum_{n=1}^{G} c_n\sigma_n$, so the measurement result of the observable $O$ can be achieved by measuring the operator $\mathrm{sign}(c_n)\sigma_n$ generated by the probability distribution $\{|c_n|/\|O\|_1\}$. When the initial state $|\psi_0\rangle$ passes through this quantum circuit, we have
\begin{align}
|\Psi_n^N\rangle=&\frac{1}{\sqrt{2}}(|0\rangle\otimes\sigma_n\cdot(U_f^{e_s})^{2k_1-k}|\psi_0\rangle+\nonumber\\
&|1\rangle\otimes(U_f^{e_s})^{2k_2-k}|\psi_0\rangle).
\end{align}
Therefore, 
\begin{align}
\langle\Psi_n^N|X|\Psi_n^N\rangle=&\mathrm{Re}(\langle\psi_0|({U_f^{e_s}}^\dag)^{2k_2-k}\cdot \mathrm{sign}(p_n)\sigma_n \cdot \nonumber\\
&(U_f^{e_s})^{2k_1-k}|\psi_0\rangle),
 \nonumber\\
\langle\Psi_n^N|Y|\Psi_n^N\rangle=&-\mathrm{Im}(\langle\psi_0|({U_f^{e_s}}^\dag)^{2k_2-k}\cdot \mathrm{sign}(p_n)\sigma_n \cdot \nonumber\\
&(U_f^{e_s})^{2k_1-k}|\psi_0\rangle). 
\end{align}
After that, the estimator $\hat{N}(n,k_1,k_2)$ can be obtained by calculating the expectation about $n$, $k_1$ and $k_2$:
\begin{align}\label{Nk1k2}
\hat{N}(n,k_1,k_2)=\left(\langle\Psi_n^N|X|\Psi_n^N\rangle-i\langle\Psi_n^N|Y|\Psi_n^N\rangle\right),
\end{align}
and then 
\begin{align}\label{N}
N=\mathbb{E}_{n,k_1,k_2}\hat{N}(n,k_1,k_2).
\end{align}
Finally, the algorithm for estimating $N(O)$ is given in Algorithm \ref{alg:N}.

\begin{figure}
  \centering
   \includegraphics[width=\linewidth]{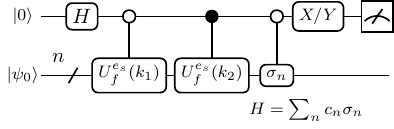}
  \caption{Quantum circuit for measuring the numerator estimator $\hat{N}_{n,k_1,k_2}$.}
  \label{fig10_circuitN}
\end{figure}

\begin{algorithm}[H]
    \caption{Estimating the numerator $N(O)$}
    \label{alg:N}
    \begin{algorithmic}[1]
    
    \vspace{0.2cm}
    \Require  An $m$-qubit observable $O=\sum_{n=1}^{G} c_n\sigma_n$; $U_f^{e_s}=\exp\left({-i(H-e_s)t}\right)$; 
    initial state $|\psi_0\rangle$ with nonzero overlap with the $j$-th eigenstate of the Hamiltonian $H$: 
    $|a_j|^2 = |\langle E_j | \psi_0 \rangle|^2\neq0$; the interval $\left[E_j^L, E_j^U\right]$ of $E_j$, and the energy-shift $e_s$ selected in $\left[E_j^L, E_j^U\right]$; 
    
    \noindent the accuracy $\epsilon$ of the Hamiltonian eigenstates $E_j$, the total failure probability bound $\delta$, and selecting the iteration times
    \[
    k=\log\left(\frac{\lVert H \rVert_1}{\epsilon|a_s|^2}\right)/{2\log\left(\frac{\lambda_s}{\lambda_t}\right)}
    \]
    according to Eq.(\ref{kk}); the properly selected sampling numbers
    \[
    n_N=O\left(\frac{\ln(\frac{1}{\delta})\cdot\lVert O \rVert_1^2}{|a_s|^4\cdot\epsilon^{-4c\cdot \log(\lambda_s)}}\right).
    \]
    
    \Ensure  Estimating the numerator $N(O)$ with failure probability $\delta_N\leqslant \delta/2$.
    \For {$q = 1$ to $n_N$}
        \State Sample an integer $n$ according to the probability distribution $p\sim\{|p_n|/\lVert O \rVert_1\}$.
        \State Sample two non-negative integers $k_1, k_2$ according to binomial distribution $B(k,1/2)$.
        \State Running the quantum circuit diagram of Fig.\ref{fig10_circuitN} to obtain the final state $|\Psi_n^N\rangle$.
        \State Implement the measurement $M$ to obtain $\langle\Psi_n|M|\Psi_n^N\rangle$, $M\in\{X,Y\}$, and is selected sequentially.
    \EndFor
    \State Calculate $\hat{N}(n,k_1,k_2)$ according to Eq.(\ref{Nk1k2}).
    \State Calculate the estimated numerator  $\overline{\hat{N}}(O)$ as an estimation of $N$:
    \[
    \overline{\hat{N}}(O)= \frac{1}{n_N} \sum_{q=1}^{n_N} \hat{N}(n,k_1,k_2).
    \]
    \end{algorithmic}
    \end{algorithm}
    

\begin{figure*}
\centering
\includegraphics[width=0.32\linewidth]{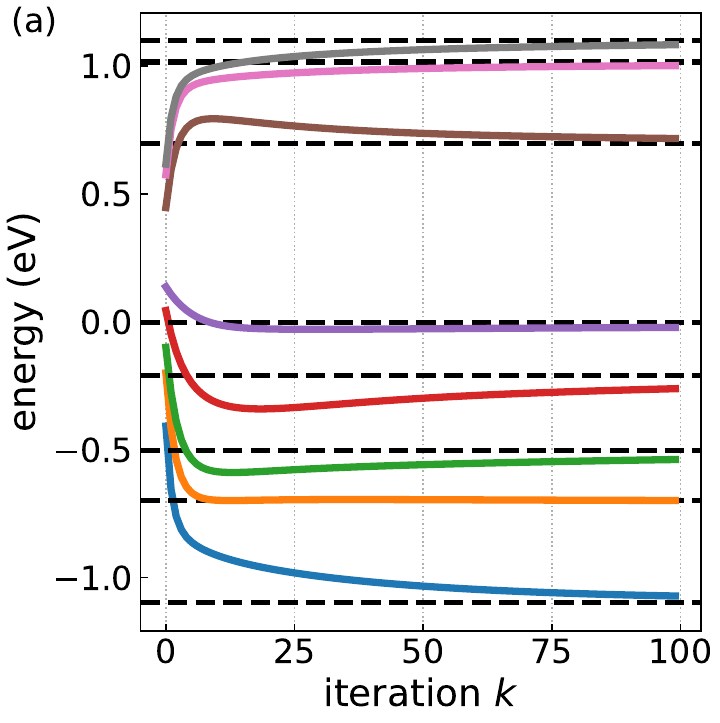}
    \includegraphics[width=0.32\linewidth]{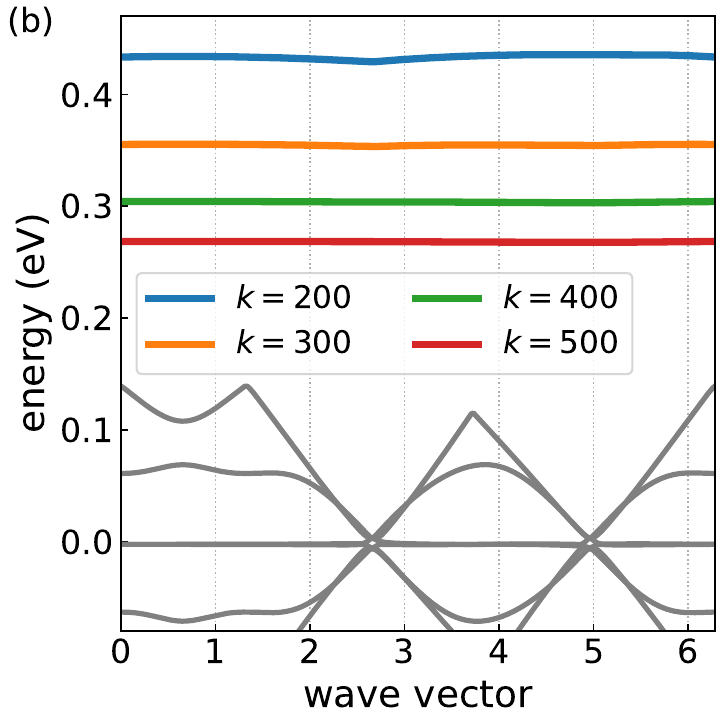}
    \includegraphics[width=0.32\linewidth]{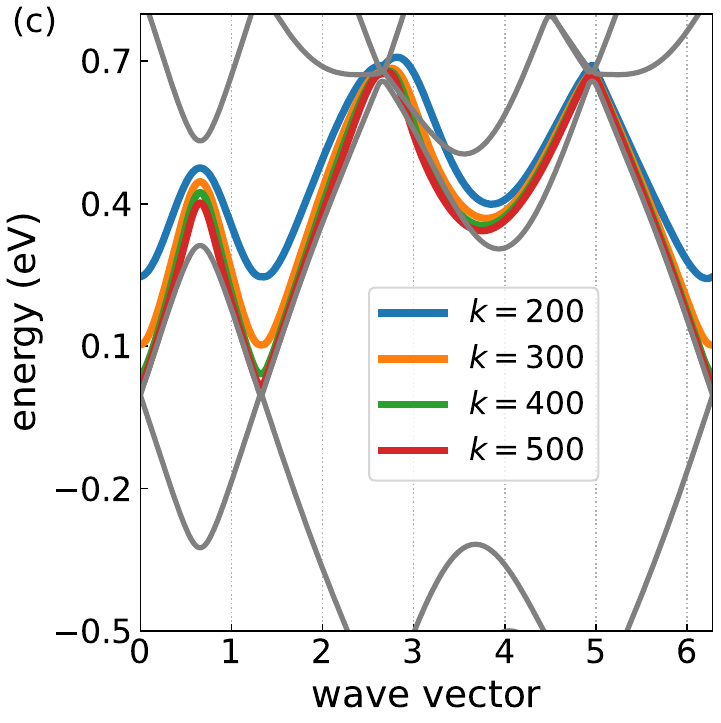}
\caption{(a) Simulation results of the iteration processes for the Bistritizer-MacDonald model~\cite{bistritzer2011moire} at a twist angle $1.05^{\circ}$ with a wave vector at $(0,0)$. The target eigenvalues are labeled by black dashed curves. The band structure of a bilayer graphene model with twist angle $1.05^{\circ}$ with evolution time $t=\pi/4$ (b) and $5^{\circ}$ with evolution time $t=\pi/20$ (c). The gray curves represent the partial band structures obtained through classical computations, while the colored solid curves represent the simulation results of our QESDS algorithm.}
\label{fig11_BM}
\end{figure*}


\section{Simulation results with the Bistritizer-MacDonald model}  \label{appenBMmodel}    

We now simulate the Bistritizer-MacDonald model of twisted bilayer graphene~\cite{bistritzer2011moire,sboychakov2015electronic,rozhkov2016electronic}, starting with the Hamiltonian at a twist angle $1.05^{\circ}$ with a wave vector at $(0,0)$. The dimension of this Hamiltonian is 196, which can be encoded into eight qubits. We select eight discontinuous energy levels $\{-1.098, -0.698, -0.502, -0.210, -0.002, 0.698, 1.016, 1.099\}$ as the simulation objects, with the initial state set as $\ket{+}^{\otimes 8}$. The evolution time is set as $t=\pi/6$, and the simulation results of the iteration processes are shown in Fig.\ref{fig11_BM}\textcolor{blue}{(a)}. Since we no longer require prior determination of results for all preceding energy levels, but instead preset estimated values for the corresponding energy levels with energy-shifts $e_{s} \in \{-1.1, -0.7, -0.5, -0.2, 0.0, 0.7, 1.0, 1.1\}$, the convergence rate for each energy level is isolated. Consequently, the convergence process for higher energy levels will be accelerated, compared to the other algorithms like the VQD~\cite{jones2019variational,higgott2019variational}, FQESS~\cite{wen2024full}, and powered-FQE~\cite{wang2024improving}.

Moreover, the presence of flat-band structures in bilayer graphene with different twist angles is of particular interest in the field of condensed matter physics~\cite{aspuru2005simulated,bistritzer2011moire,yin2022topological}. Flat bands are regions in momentum space where the energy exhibits minimal variation with respect to momentum, often associated with strong correlation phenomena. We use our algorithm to investigate this phenomenon, selecting bilayer graphene models with twist angles of $1.05^{\circ}$ and  $5^{\circ}$ and setting $e_s = 0$. To avoid cancellations of positive and negative contributions arising from the symmetry of the band structure, we first measure the expectation value of the square of the Hamiltonian and then take its square root, $ \vert E_i \vert = \sqrt{\langle E_i |H^2 |E_i \rangle} $. The simulation results for different iterations are shown in Fig.\ref{fig11_BM}\textcolor{blue}{(b,c)}. We can conclude from the results that a flat band exists in the bilayer graphene model with a twist angle of $1.05^{\circ}$, but not with a twist angle of $5^{\circ}$. This prediction is consistent with theoretical expectations. Compared to other quantum algorithms, the QESDS algorithm can directly assess the presence of a flat band without needing to compute lower-energy bands.

\newpage

\bibliography{QESDS.bib}


\end{document}